\documentclass[aps,prl,reprint,twocolumn,showpacs,amsmath,amssymb,amsfonts,10pt,nofootinbib]{revtex4-2}

\usepackage[T1]{fontenc}
\usepackage{graphicx}
\usepackage{enumitem}
\usepackage{amsmath}
\usepackage{amsfonts}
\usepackage{amssymb}
\usepackage{mwe}
\usepackage{color}
\usepackage[dvipsnames]{xcolor}
\usepackage{bm}
\usepackage{soul}
\usepackage[normalem]{ulem}

\usepackage[colorlinks,bookmarks=false,urlcolor=blue,citecolor=blue,linkcolor=blue]{hyperref}
\usepackage[capitalise]{cleveref}
\usepackage{comment}

\usepackage[compat=1.1.0]{tikz-feynman}
\usepackage[compat=1.1.0]{tikz-feynhand}
\tikzset{ultra thick/.style={line width=2pt}}
\usepackage{adjustbox}

\newcommand{\setsmalltitle}[1]{{\textbf{\textit{#1}.}---}}

\newcommand{\ket}[1]{|{#1}\rangle}

\newcommand{\veck}{\mathbf k}

\newcommand{\ee}{\mathrm{e}} 

\DeclareRobustCommand{\circleletter}[1]{%
  \raisebox{-1.5pt}{\tikz[baseline={(char.south)}]{%
    \node[shape=circle,draw,inner sep=0.5pt] (char) {#1};%
  }%
  }%
}
\begin{document}

\title{Mass-gap description of heavy impurities in Fermi gases}

\author{Xin Chen$^*$}
\author{Eugen Dizer$^*$}
\author{Emilio Ramos Rodr\'iguez}
\author{Richard Schmidt}
\affiliation{$ $Institut f\"ur Theoretische Physik, Universit\"at Heidelberg, D-69120 Heidelberg, Germany}

\begin{abstract}
We present a unified theory that connects the quasiparticle picture of Fermi polarons for mobile impurities to the Anderson orthogonality catastrophe for static impurities. By operator reordering of the underlying many-body Hamiltonian, we obtain a modified fermionic dispersion relation that features a recoil-induced energy gap, which we call the `mass gap'. We show that the resulting mean-field Hamiltonian exhibits an in-gap state for finite impurity mass, which takes a key role in Fermi polaron and molecule formation.
We identify the mass gap as the microscopic origin of the quasiparticle weight of Fermi polarons and derive a power-law scaling of the weight with the impurity-to-fermion mass ratio. The associated in-gap state is shown to give rise to the emergence of the polaron-to-molecule transition away from the limiting case of the Anderson orthogonality catastrophe in which the transition is absent.
\end{abstract}

\maketitle

\def\thefootnote{*}\footnotetext{These authors have contributed equally to this work.}\def\thefootnote{\arabic{footnote}}

In 1967, Anderson proved that the ground state of a system of $N$ fermions interacting with a static impurity of infinite mass is orthogonal to the ground state of the system without the impurity, when ${N\rightarrow\infty}$~\cite{Anderson:1967}. This phenomenon is known as the orthogonality catastrophe (OC) and represents a fundamentally non-perturbative effect~\cite{Nozieres:1969}. As a result, path integral methods~\cite{Prokof'ev:2008,Prokof'ev:2008_2,Tilman2011,Kroiss2015,Hu2024,Hui2024SpecPRL}, or variational and diagrammatic approaches using perturbative expansions~\cite{Chevy2006,Combescot2007,Combescot2008,Punk:2009PolMol,Mora:2009,Cui2010,Bruun:2010PolMol,Trefzger2012,Schmidt2012,Liu:2019ThermoVA,Rana2021} fail to capture the OC. On the other hand, such methods have been demonstrated to work remarkably well for the case of mobile impurities of finite mass, which is governed by the formation of Fermi polarons featuring a finite quasiparticle weight~\cite{Massignan:2014,Schmidt2018,Scazza2022,massignan2025polarons}. 

Fermi polaron formation proceeds similarly to the paradigm considered by Landau and Pekar~\cite{Landau:1933,Landau:1948ijj}, where mobile impurities interact with a bosonic bath to form polaron quasiparticles \cite{Grusdt_2025} characterized by an effective mass $m^*$, polaron energy $E_\text{pol}$, quasiparticle weight $Z$ and lifetime $\tau$ \cite{Mona:2006}. In the case of Fermi polarons, quasiparticle properties have been observed first in ultracold atoms~\cite{Schirotzek:2009exp,Kohstall2012,Thomas2012,Cetina2016,Yan2019,Adlong:2020,Hu2022RamanSpecFermiPolaron,Baroni2024,Baroni2024FP,Vivanco2025} and later in two-dimensional materials~\cite{Sidler:2017,Efimkin:2018XPol,Chervy:2020Polariton,imamoglu2021exciton,Rana2021,Koksal2021,Xiaoqin2023}. Additionally, for a finite impurity mass, the fermionic system is believed to feature a sharp polaron-to-molecule transition~\cite{Prokof'ev:2008,Prokof'ev:2008_2,Punk:2009PolMol,Bruun:2010PolMol,Parish:2011_2DPolMol,Tilman2011,Ness:2020PolMol,Cui2020,Cui:2021PolMol,Parish2021}, which is absent in the OC scenario.  

Connecting the quasiparticle picture of Fermi polarons with Anderson’s orthogonality catastrophe 
presents an outstanding challenge \cite{Pimenov2017,Pimenov:2018,Gievers2025}. Due to the non-perturbative nature of the OC and the few-body character of the polaron-to-molecule transition, a unified description has remained elusive \cite{Schmidt2018,massignan2025polarons}.

In this letter, we present a theory for finite-mass impurities in Fermi gases that captures the Anderson OC,  the quasiparticle description of Fermi polarons and the polaron-to-molecule transition in one unified model. Utilizing canonical transformations, operator reordering and exact diagonalization, our theory provides a description of how quasiparticle behavior emerges from the OC, and how the existence of the polaron-to-molecule transition can be understood in terms of an effective in-gap state emerging in the model. While we focus here on the case of contact interactions in three dimensions at zero temperature,  our theory can be applied to arbitrary interaction potentials, dimensions and temperatures.

\setsmalltitle{Model} The Hamiltonian for an impurity of mass $M$ interacting with a Fermi sea of particles of mass $m$ is
\begin{align} \label{eq:polaron-hamiltonian}
    \hat{H} =\, &\sum_{\mathbf{k}} \frac{\mathbf{k}^2}{2M}\, \hat{d}^{\dagger}_{\mathbf{k}}\hat{d}_{\mathbf{k}} + \sum_{\mathbf{k}} \frac{\mathbf{k}^2}{2m}\, \hat{c}^{\dagger}_{\mathbf{k}}\hat{c}_{\mathbf{k}} \notag \\   
    &
    + \frac{g}{\mathcal{V}} \sum_{\mathbf{k}',\mathbf{k},\mathbf{q}} \hat{d}^{\dagger}_{\mathbf{k}'+\mathbf{q}}\hat{d}_{\mathbf{k}'} \hat{c}^{\dagger}_{\mathbf{k}-\mathbf{q}}\hat{c}_{\mathbf{k}} \,.
\end{align}
The first two terms describe the kinetic energy of the impurity ($\hat{d}^{\dagger}_{\mathbf{k}}$) and host fermions ($\hat{c}^{\dagger}_{\mathbf{k}}$), respectively. The third term represents a contact interaction between the two species with bare coupling constant $g$ and quantization volume $\mathcal{V}$. The value of $g$ is related to the physical scattering length $a$ via ${1/g = m_r/(2\pi a) - m_r \Lambda/\pi^2}$, where ${m_r = mM/(m+M)}$ is the reduced mass and $\Lambda$ is a momentum cutoff.

As a first step, we decouple the degrees of freedom of the impurity from the bath by performing a canonical transformation. To this end, the Hamiltonian $\hat{H}$ in Eq.~\eqref{eq:polaron-hamiltonian} is expressed in terms of the impurity's single-particle position and momentum operators $\hat{\mathbf{r}}$ and  $\hat{\mathbf{P}}$, respectively. The interaction term then couples the impurity and Fermi gas through an $\hat{\mathbf{r}}$-dependent phase ${\sim\ee^{i {\mathbf{q}}\cdot \hat{\mathbf{r}}}}$. Applying the unitary Lee-Low-Pines (LLP) transformation ${\hat{U} = \ee^{i\hat{\mathbf{r}}\cdot\sum_{\mathbf{k}} \mathbf{k}\, \hat{c}^{\dagger}_{\mathbf{k}}\hat{c}_{\mathbf{k}}}}$, i.e., ${\hat{\mathcal{H}}=\hat{U}\hat{H}\hat{U}^{-1}}$, removes the operator-valued phase from the interaction term, effectively decoupling the impurity and bath Hilbert spaces~\cite{Lee:1953}. However, since $\hat{U}\hat{\mathbf{P}}\hat{U}^{-1}=\hat{\mathbf{P}}-\sum_{\mathbf{k}} \mathbf{k}\, \hat{c}^{\dagger}_{\mathbf{k}}\hat{c}_{\mathbf{k}}$ this transformation also acts on the kinetic energy operator of the impurity, introducing an effective interaction between Fermi sea particles, ${\hat{\mathcal{H}}_{\text{int}}=\frac{1}{2M}\sum_{\mathbf{kk}'}(\veck \cdot \veck')\hat{c}^{\dagger}_{\mathbf{k}}\hat{c}^{\dagger}_{\mathbf{k}'}\hat{c}_{\mathbf{k}'}\hat{c}_{\mathbf{k}}}$.

\begin{figure}[t!]
    \centering
    \includegraphics[width=0.49\textwidth]{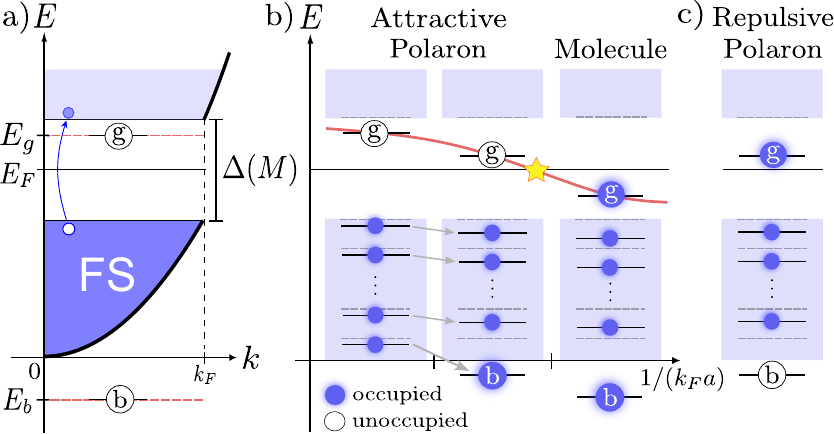}
    \caption{\textbf{Mass-gap model.} a) Sketch of the modified dispersion relation $E_{\mathbf{k}}$ and opening of the mass gap $\Delta(M)$. The in-gap state \circleletter{g} and bound state \circleletter{b} are depicted with a red dashed line. A particle-hole excitation from the lower to the upper band is shown in blue. b) As the interaction strength increases, the single-particle energies (solid lines) decrease compared to the non-interacting levels (dashed lines), and the in-gap state energy $E_g$ moves from the edge of the upper band to the lower band edge. The polaron-to-molecule transition (depicted as  yellow star) occurs at the critical interaction strength $1/(k_F a_c)$, when the in-gap state crosses the Fermi energy $E_F$. The attractive polaron is described by occupying the lowest energy states, including the bound state \circleletter{b}. The molecule is obtained by the additional occupation of the in-gap state \circleletter{g}. c) The repulsive polaron corresponds to the excited configuration occupying the in-gap state instead of the bound state.}
    \label{fig:modified-dispersion-relation}
\end{figure}

We proceed by splitting the momentum summations of $\hat{\mathcal{H}}_{\text{int}}$ into contributions below and above the Fermi momentum $k_F$, that is,  $\sum_{\mathbf{k}}=\sum_{|\mathbf{q}|< k_F}+\sum_{|\mathbf{k}|>k_F}$. This procedure yields a new Hamiltonian
which, for a system of total momentum $P=0$, is given by (for details see the Supplemental Material~\cite{Supplemental})
\begin{align} \label{eq:gapped-Hamiltonian}
    \hat{\mathcal{H}}= E_0 + \sum_{\mathbf{k}} E_{\mathbf{k}} \hat{c}^{\dagger}_{\mathbf{k}}\hat{c}_{\mathbf{k}} + \frac{g}{\mathcal{V}} \sum_{\mathbf{k},\mathbf{k}'} \hat{c}^{\dagger}_{\mathbf{k}}\hat{c}_{\mathbf{k}'} + \hat{\mathcal{H}}_{\mathrm{int}}^\mathrm{LLP} \,.
\end{align}
We term Eq.~\eqref{eq:gapped-Hamiltonian} the \textit{mass-gap model}. It is formally equivalent to the original Hamiltonian Eq.~\eqref{eq:polaron-hamiltonian} but, as we will demonstrate in the following, allows to address the connection between the Fermi polaron problem and the OC from a new angle. 
In the mass-gap model~\eqref{eq:gapped-Hamiltonian}, $E_0 = \sum_{|\mathbf{q}|< k_F} \mathbf{q}^2/(2M)$ is a zero-point energy and $\hat{\mathcal{H}}_{\mathrm{int}}^\text{LLP}$ takes the form
\begin{align}
\hat{\mathcal{H}}_{\mathrm{int}}^\text{LLP}=~&\frac{1}{2M} \sum_{|\mathbf{q}|,|\mathbf{q}'|< k_F} \left(\mathbf{q}\cdot\mathbf{q}'\right) \hat{c}_{\mathbf{q}}\hat{c}_{\mathbf{q}'} \hat{c}^{\dagger}_{\mathbf{q}'}\hat{c}^{\dagger}_{\mathbf{q}}\nonumber\\
&-\frac{1}{M} \sum_{|\mathbf{q}|<k_F<|\mathbf{k}|} \left(\mathbf{k}\cdot\mathbf{q}\right) \hat{c}^{\dagger}_{\mathbf{k}}\hat{c}_{\mathbf{q}}\hat{c}^{\dagger}_{\mathbf{q}}\hat{c}_{\mathbf{k}}\nonumber\\
&+\frac{1}{2M} \sum_{|\mathbf{k}|,|\mathbf{k}'|>k_F} \left(\mathbf{k}\cdot\mathbf{k}'\right) \hat{c}^{\dagger}_{\mathbf{k}}\hat{c}^{\dagger}_{\mathbf{k}'}\hat{c}_{\mathbf{k}'}\hat{c}_{\mathbf{k}}\, .
\end{align}
In this way, the interaction terms are reordered so that they vanish when acting on the Fermi sea ${\ket{\text{FS}}=\prod_{|\mathbf{q}|<k_F}\hat{c}^\dagger_{\mathbf{q}}\ket{0}}$.

\begin{figure}[t!]
    \centering
    \includegraphics[width=0.48\textwidth]{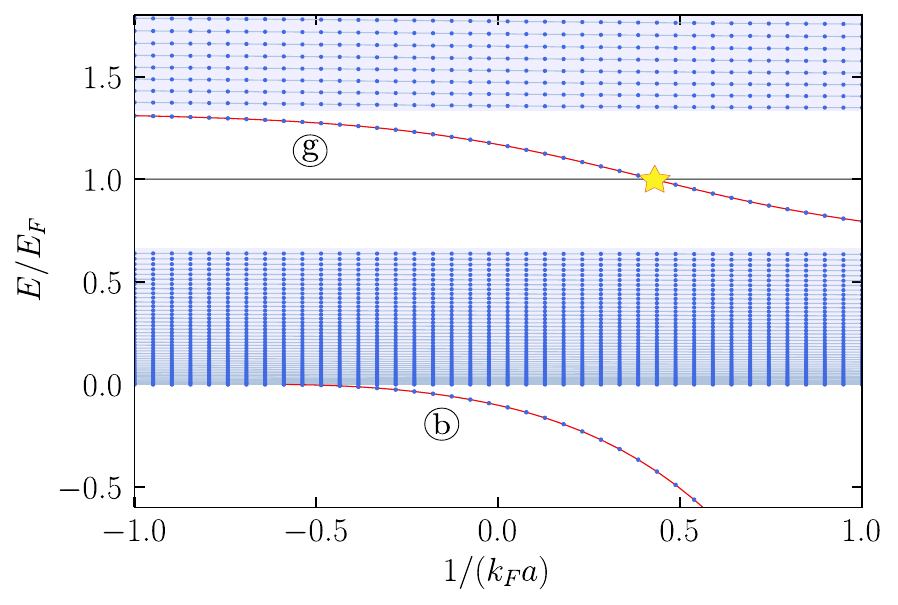}
    \caption{\textbf{Exact energy spectrum.} Energy spectrum     of the gapped Hamiltonian $\hat{\cal H}_\text{quad}$ obtained from exact diagonalization at finite mass ratio $M/m=3$ as function of interaction strength $1/(k_Fa)$. Due to the gapped dispersion relation $E_{\mathbf{k}}$, the bound state \circleletter{b} appears already at a  negative scattering length. The in-gap state \circleletter{g} crosses the Fermi level $E_F$ (solid line) at positive scattering length, signaling the polaron-to-molecule transition (yellow star).}
    \label{fig:numerical-spectrum-plot}
\end{figure}

Importantly, the  Hamiltonian~\eqref{eq:gapped-Hamiltonian} contains a modified dispersion relation $E_{\mathbf{k}}$ of the fermions given by
\begin{align} \label{eq:modified-dispersion}
    E_{\mathbf{k}} =
    \begin{cases}
        \frac{\mathbf{k}^2}{2m} - \frac{\mathbf{k}^2}{2M} & \quad\mathrm{for}\quad |\mathbf{k}|<k_F \,, \\[1ex]
        \frac{\mathbf{k}^2}{2m} + \frac{\mathbf{k}^2}{2M} & \quad\mathrm{for}\quad |\mathbf{k}|>k_F \,.
    \end{cases}
\end{align}
This dispersion relation, previously described by Kain and Ling~\cite{Kain:2017HF}, features an energy gap at the Fermi momentum $k_F$, which is related to the recoil energy of the impurity (for an illustration see \cref{fig:modified-dispersion-relation}(a)). We will see that this impurity-mass induced gap has important implications for the Fermi polaron problem, and serves as a regulator for the orthogonality catastrophe. From now on, we will call it the \textit{mass gap}.

\setsmalltitle{Mass-gap description} The mass gap around the Fermi level,
\begin{align} \label{eq:mass-gap}
    \Delta(M) = \frac{k^2_F}{M} \,,
\end{align}
depends on the density of the Fermi sea through $k_F$. As a consequence, already the {\it single-particle} eigenstates $|\nu\rangle$ of the quadratic `mean-field' Hamiltonian ${\hat{\cal H}_\text{quad}=\hat{\cal H}-\hat{\cal H}_\text{int}^\text{LLP}-E_0}$, i.e.,
\begin{align} \label{eq:quadratic-Hamiltonian}
    \hat{\cal H}_\text{quad}=  \sum_{\mathbf{k}} E_{\mathbf{k}} \hat{c}^{\dagger}_{\mathbf{k}}\hat{c}_{\mathbf{k}} + \frac{g}{\mathcal{V}} \sum_{\mathbf{k},\mathbf{k}'} \hat{c}^{\dagger}_{\mathbf{k}}\hat{c}_{\mathbf{k}'} \,,
\end{align}
carry information about the many-body physics.

From the interacting orbitals $|\nu\rangle$, one can construct the  exact many-body ground state $| \widetilde{\text{FS}}\rangle$ of $\hat{\cal H}_\text{quad}$ as a Slater determinant. Importantly, as a  consequence of the reordering of the LLP interaction, it follows that $\langle \widetilde{\text{FS}}|\hat{\cal H}_\text{int}^\text{LLP}|\widetilde{\text{FS}}\rangle=0$ for s-wave interactions.  Moreover, using $\langle \widetilde{\text{FS}}|\hat{\cal H}|\widetilde{\text{FS}}\rangle = \langle \widetilde{\text{FS}}|\hat{\cal H}_\text{quad}|\widetilde{\text{FS}}\rangle+E_0$, it follows that the ground-state energy of $\hat{\cal H}_\text{quad}$ sets an upper variational bound for the ground-state energy of the full Hamiltonian $\hat{\cal H}$, and thus also $\hat H$. A  detailed discussion of the  relation between the ground state of $\hat{\cal H}_\text{quad}$ and $\hat{\cal H}$ is provided in \cite{Supplemental}. 

\begin{figure}[t]
    \centering
    \includegraphics[width=0.49\textwidth]{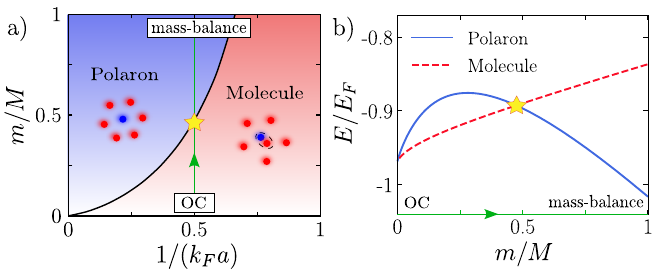}
    \caption{\textbf{Polaron-to-molecule transition from the OC.} a) Phase diagram of the impurity problem in 3D as function of the inverse mass ratio $m/M$ and interaction strength $1/(k_Fa)$ as predicted by our theory. b) Polaron and molecule energy as a function of  $m/M$ for a fixed interaction strength $1/(k_Fa)=0.5$. The polaron-to-molecule transition is highlighted with a yellow star.}
    \label{fig:polaron-to-molecule-transition}
\end{figure}

The spectrum of $\hat{\cal H}_\text{quad}$, calculated using exact diagonalization, is shown in \cref{fig:numerical-spectrum-plot} as a function of the interaction strength $1/(k_F a)$. For any finite mass ratio $M/m$ the spectrum exhibits an in-gap state \circleletter{g}, see Fig.~\ref{fig:modified-dispersion-relation}(a). Moreover, for sufficiently strong attraction, a bound state \circleletter{b} appears. The energies of both the state \circleletter{g} and \circleletter{b} are encoded in the poles of the \textit{in-medium} $T$-matrix, $\tilde{T}(E)=(1/g - 1/{\cal V}\sum_{\mathbf{k}} 1/(E-E_\mathbf{k}))^{-1}$, which, compared to the standard two-body problem, is modified by the gapped dispersion relation $E_{\bf k}$. Due to the mass gap, the bound state \circleletter{b} already emerges at negative scattering length.
A sign analysis of $\tilde{T}(E)$ within the energy gap confirms the universal existence of the interaction-induced in-gap state for all values of the coupling strength $g$. An analytical calculation of $\tilde{T}(E)$ for  s-wave contact interactions is given in \cite{Supplemental}. 

The many-body properties of $\hat{\cal H}_\text{quad}$ can be understood based on the single-particle eigenenergies $\omega_\nu$ and the related orbitals $|\nu\rangle$, which are occupied by a finite number of fermions. As illustrated in \cref{fig:modified-dispersion-relation}(b,c), three  fermion configurations can be distinguished, depending on whether the in-gap state \circleletter{g} and the bound state \circleletter{b} are occupied or not. These individual combinations are, in fact, related to the three major excitations of the Fermi polaron problem: the attractive polaron and  molecule (\cref{fig:modified-dispersion-relation}(b)), and the repulsive polaron (\cref{fig:modified-dispersion-relation}(c)). 

The  attractive Fermi polaron is described by the configuration in which all $N$ particles occupy the  lowest energy states, excluding the in-gap state. The polaron energy $E_\text{pol}$ is then given by the sum of all energy level shifts resulting from the presence of the impurity potential, including the bound state energy $E_b$, if \circleletter{b} is present. In the thermodynamic limit, the polaron energy is  given by a modification of Fumi's theorem~\cite{Fumi:1955} as sum over all scattering phase shifts $\tilde{\delta}(E)$, i.e., $E_{\mathrm{pol}} = -\int_0^{E_F} \tilde{\delta}(E)/\pi\, dE + E_b$, where the scattering phase $\tilde{\delta}(E)$ is determined by the \textit{in-medium} $T$-matrix $\tilde{T}(E)$. It thus contains knowledge about the many-body physics through the density-dependent mass gap $\Delta(M)$, and hence differs from the two-body scattering phase $\delta(E)$ in vacuum.

The dressed molecular state, sometimes referred to as the `molaron' \cite{Mora:2009,Cui:2021PolMol,Oriana:2024Eject,massignan2025polarons}, is represented by occupying the in-gap state \circleletter{g} in addition to the bound state \circleletter{b} (Fig.~\ref{fig:modified-dispersion-relation}(c)). In order to conserve the overall particle number, the number of states in the lower band must now be $N-1$, leading to a shift in the energy of the state by the Fermi energy $E_F$. The molecular energy is thus given by $E_{\mathrm{mol}}=E_{\mathrm{pol}}+E_g-E_F$, where $E_g$ is the energy of the in-gap state. A detailed analysis of the polaron and molecule energy as function of $1/(k_F a)$ and inverse mass ratio $m/M$ can be found in~\cite{Supplemental}, leading to the phase diagram of the polaron-to-molecule transition  discussed below (Fig.~\ref{fig:polaron-to-molecule-transition}). The expression for the polaron energy $E_\text{pol}$ using Fumi's theorem becomes exact in the infinite-mass limit, where it is equal to the molecular energy $E_\mathrm{mol}$. Finite-mass corrections only enter at second order of perturbation theory in $\hat{\cal H}_\text{int}^\text{LLP}$ (note that, within the mass-gap description, perturbation theory in $\hat{\cal H}_\text{int}^\text{LLP}$ is well defined).

The repulsive polaron corresponds to the occupation of the in-gap state \circleletter{g}, while leaving the bound state \circleletter{b} empty. The energy of the repulsive polaron is then given by $E_{\mathrm{rep}}=E_{\mathrm{pol}}+E_g-E_b$. This means that the repulsive polaron only exists if the bound state is present. As we have seen, in the presence of the Fermi bath, the bound state \circleletter{b} already appears at a negative scattering length which agrees with findings using non-self-consistent $T$-matrix  and various variational approaches \cite{Tilman2011,Massignan2011,Ngampruetikorn_2012}, further demonstrating how the mass-gap description can capture key predictions, both in the finite and infinite-mass limit. 

\setsmalltitle{Polaron-to-molecule transition} One key feature of the mass-gap model is that it provides a natural explanation for the emergence of the polaron-to-molecule transition  as the impurities become mobile. In this description, the new in-gap state \circleletter{g} plays a central role. Specifically, at a critical interaction strength, $1/(k_Fa_c)$, the in-gap state crosses the Fermi level $E_F$ and becomes energetically favorable to occupy. At this point, the molecule becomes the new ground state of the system; see \cref{fig:modified-dispersion-relation}(b) for an illustration and \cref{fig:polaron-to-molecule-transition} for numerical results. Since the energies cross with a discontinuous derivative, the ground state energy exhibits the characteristics of a first-order phase transition. The value of $1/(k_Fa_c)$ depends on the mass ratio, leading to the `phase diagram' shown in \cref{fig:polaron-to-molecule-transition}(a). In general, we find that the molecule is favorable for larger mass ratios. Moreover, approaching the infinite-mass limit, the transition occurs at unitarity, $1/(k_Fa) = 0$.

For a given mass ratio $M/m$, the difference between the polaron and molecule energy is bounded by ${|E_{\mathrm{pol}}-E_{\mathrm{mol}}|<\Delta(M)/2}$. In the limit  $1/(k_Fa)\rightarrow\pm\infty$, the separation between the polaron and molecule energy is exactly $\Delta(M)/2$. Thus, by measuring this energy difference, the effect of the mass gap $\Delta(M)$ can be observed in experiment. In the limit $M\rightarrow\infty$, the mass gap closes and the in-gap state vanishes. Here the polaron and molecule states become equivalent, and the transition ceases to exist. The remaining state no longer exhibits quasiparticle properties and the ground state becomes orthogonal to the non-interacting Fermi sea. The ability of the mass-gap description to capture the merging of the polaron and molecule  in the OC limit is in stark contrast to the usual variational treatments~\cite{Punk:2009PolMol,Bruun:2010PolMol,Parish:2011_2DPolMol}, where they always remain distinct states. 

Remarkably, despite the simplicity of the effective mean-field approach which only involves $\hat{\cal H}_\text{quad}$, we obtain an accurate prediction for the polaron-to-molecule transition even as $m/M\to 1$, where higher-order quantum fluctuations induced by $\hat{\cal H}_\text{int}^\text{LLP}$ are expected to become relevant. Specifically, even for the unfavorable case $m/M=1$, we obtain a critical interaction strength of $1/(k_Fa_c)=0.69$, which compares well with the diagrammatic Monte Carlo result of
$1/(k_Fa_c)=0.90$~\cite{Prokof'ev:2008,Prokof'ev:2008_2}, while the non-selfconsistent T-matrix approach yields a value of $1/(k_Fa_c) = 1.27$~\cite{Punk:2009PolMol}. Given that the mass-balanced limit is at the edge of the mean-field theory’s applicability, the level of agreement is quite remarkable. 

\setsmalltitle{Quasiparticle weight}
The second important feature of the mass-gap description is the ability to explain the generation of a quasiparticle weight $Z$ for impurities of finite mass $M$. The presence of a finite mass gap $\Delta(M)>0$ is central,  as it serves as  a regulator for the low-energy excitations  associated with the Anderson orthogonality catastrophe, where $Z=0$. By suppressing these low-energy particle-hole excitations, $\Delta(M)$ ensures a finite overlap between interacting and non-interacting  ground states, thus establishing a unified framework  connecting OC physics with the quasiparticle picture. This suppression also provides a natural explanation for the remarkable efficiency of the variational Chevy ansatz~\cite{Chevy2006} for light impurity masses, where each particle-hole excitation comes with an energy cost given by the gap. 

\begin{figure}[t]
    \centering
    \includegraphics[width=0.49\textwidth]{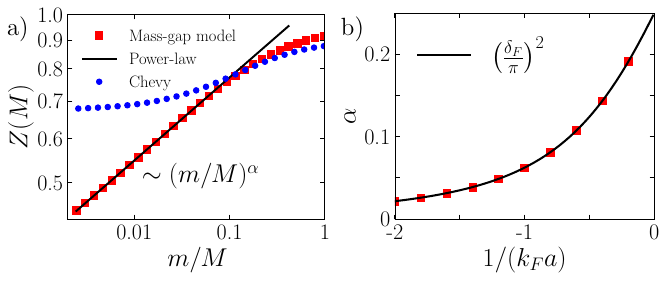}
    \caption{\textbf{Emergence of the quasiparticle weight.}
    a) Quasiparticle weight $Z$ of the attractive polaron as function of the mass ratio $m/M$. Red symbols represent the mean-field prediction of the mass-gap model. The solid line corresponds to a power-law with exponent $\alpha = (\delta_F/\pi)^2$. Blue symbols show results from the Chevy ansatz~\cite{Chevy2006}, which fails to capture the OC. b) Numerically calculated exponent $\alpha$ of the power-law decay $Z(M)\sim (m/M)^{\alpha}$ (red symbols) as function of $1/(k_Fa)$ compared to the analytical expression $(\delta_F/\pi)^2$ (solid line).}
    \label{fig:quasiparticle-weight}
\end{figure}

Quantitatively, the quasiparticle weight is obtained in the mass-gap model by the calculation of the overlap $Z=|\langle\mathrm{FS}|\widetilde{\text{FS}}\rangle|^2$, where $|\widetilde{\text{FS}}\rangle$ and $|\mathrm{FS}\rangle$ denote the ground state of $\hat{\cal H}_\text{quad}$ and the non-interacting Fermi sea, respectively. The overlap is readily calculated by a Slater determinant~\cite{Anderson:1967},
\begin{align}
    Z = \det
    \begin{bmatrix}
        \langle n=1|\nu=1 \rangle & \cdots & \langle n=1|\nu=N \rangle \\
        \vdots &  \ddots & \vdots  \\
        \langle n=N|\nu=1 \rangle & \cdots & \langle n= N|\nu =N\rangle
    \end{bmatrix}^2 \,.
\end{align}
Here, $N$ is the highest occupied single-particle state, while $n$ and $\nu$ label the quantum numbers of the single-particle orbitals of the non-interacting Hamiltonian and $\hat{\cal H}_\text{quad}$, respectively. The quasiparticle weight of the attractive polaron is shown in \cref{fig:quasiparticle-weight}, where we also show a direct comparison to the variational Chevy ansatz \cite{Chevy2006}.
The numerical result (red symbols) shows  a universal power-law scaling (solid line) of the quasiparticle weight $Z(M)$ with respect to the mass ratio $M/m$, 
\begin{align}
    \label{eq:quasiparticle-weight}
    Z(M) \sim (M/m)^{-\alpha} \,, \qquad \alpha = \left(\frac{\delta_F}{\pi}\right)^2 \,,
\end{align}
where $\delta_F\equiv\delta(k_F)$ represents the scattering phase shift of the infinite-mass impurity at $k_F$, obtained by ${\cot(\delta(k))=-1/(ka)}$. Eq.~\eqref{eq:quasiparticle-weight} applies for large enough  mass ratios, in absence of the bound state \circleletter{b}. In its presence, i.e., for positive scattering length $a$, the power-law coefficient $\alpha$ is modified by an additional factor \cite{Knap:2012OC,Schmidt2018}. For more details, see the Supplemental Material~\cite{Supplemental}.

The scaling in Eq.~\eqref{eq:quasiparticle-weight} is reminiscent of the system-size power-law scaling $Z\sim N^{-\alpha}$ of the infinite-mass impurity \cite{Anderson:1967}. In this analogy,  $\Delta(M)$ effectively replaces the conventional infrared cutoff $E_F/N$ involved in the calculation for static impurities in a finite system by Hamann~\cite{Hamann1971}, in accordance with an argument given by Rosch~\cite{Rosch:1999}. A similar calculation for the molecule ground state shows that its $Z$ factor vanishes for all mass ratios as $N\rightarrow\infty$, attesting to the role of the molecule as a dark state in the Fermi polaron problem \cite{Supplemental}. 

Inspection of Fig.~\ref{fig:quasiparticle-weight} demonstrates another interesting aspect of our prediction: while the quasiparticle weight does decay to zero for large mass ratios $M/m$, this decay is exceedingly slow, even at large interaction strengths. E.g., for  $1/(k_Fa) = -0.4$, we find  for a ratio of $m/M=6/133 = 0.045 $, corresponding to a Li-Cs mixture, that the quasiparticle weight only decayed to roughly 50\% of its value at mass balance. Moreover, our results show that the onset of  OC power-law scaling is in range of current experimental setups. The comparison to Chevy's ansatz shows  that cold-atom based quantum simulators can serve as a clear future benchmark discerning predictions of various many-body theories.

\setsmalltitle{Conclusion} We have presented a unified theoretical framework for mobile impurities in a Fermi sea that bridges Anderson's orthogonality catastrophe with the quasiparticle picture of Fermi polarons. We analyzed the mass-gap model on a mean-field level that smoothly connects to the exact Hamiltonian of the infinite-mass impurity. A crucial aspect of our theory is the existence of an in-gap state due to the mass gap generated by the recoil of the impurity.  This phenomenon helps to explain the microscopic origin of the polaron-to-molecule transition and the scaling of the quasiparticle weight with the impurity mass. The analysis can be improved by including the recoil-induced interactions $\mathcal{H}_\text{int}^\text{LLP}$. This opens up a  path to analyze the Fermi polaron problem from a new angle using established quantum field theory, diagrammatic Monte-Carlo or variational techniques.  We note that the mechanism presented in this letter can be applied directly to other population-imbalanced fermionic systems, irrespective of spatial dimensionality or details of the underlying interaction potentials. As such, further analysis of the presented model may allow to address outstanding questions on Fermi polarons, including their dynamic or thermodynamic properties.

\setsmalltitle{Acknowledgements} We thank A.~Christianen, M.~Gievers and P.~Hill for helpful discussions. Funded by the DFG (German Research Foundation) – Project-ID 273811115 – SFB 1225 ISOQUANT, and under Germany's Excellence Strategy EXC 2181/1 - 390900948 (the Heidelberg STRUCTURES Excellence Cluster).

\appendix
\clearpage
\newpage
\begin{widetext}
\begin{center}
\textbf{\large Supplemental Material for \\``Mass-gap description of heavy impurities in Fermi gases''}\label{Sec:Supp}
\end{center}

\begin{center}
Xin Chen$^*$, Eugen Dizer$^*$, Emilio Ramos Rodr\'iguez, Richard Schmidt 
\end{center}
	
\setcounter{equation}{0}
\setcounter{figure}{0}
\setcounter{table}{0}
\setcounter{page}{1}
\makeatletter
\renewcommand{\theHequation}{S\arabic{equation}}
\renewcommand{\theHfigure}{S\arabic{figure}}
\renewcommand{\theHtable}{S\arabic{table}}
\renewcommand{\theequation}{S\arabic{equation}}
\renewcommand{\thefigure}{S\arabic{figure}}
\renewcommand{\thetable}{S\arabic{table}}

\def\thefootnote{*}\footnotetext{These authors have contributed equally to this work.}\def\thefootnote{\arabic{footnote}}

This Supplemental Material provides details supporting the work ``Mass-gap description of heavy impurities in Fermi gases''. Section \hyperref[sec:LLP-transformation]{I} reviews the Lee-Low-Pines transformation.  Section \hyperref[sec:infrared-divergencies]{II} explains the theoretical challenge to address infrared singularities which motivate the introduction of the mass-gap description. Section \hyperref[sec:gapped-dispersion]{III} provides a detailed derivation of \cref{eq:gapped-Hamiltonian} from the main text by appropriate operator reordering of the Hamiltonian. Section \hyperref[sec:exact-solution]{IV} shows the analytical $T$-matrix calculation for the scattering of the impurity with the  fermions with gapped dispersion. This section also provides a  discussion on why the ground-state energy of the quadratic Hamiltonian is an upper bound for the ground-state energy of the total Hamiltonian. Section \hyperref[sec:polaron-to-molecule-2]{V} discusses the calculation of the molecular state energy, with particular focus on the difference of the Fermi energy compared to the polaron state encountered in this case. Moreover, a detailed `phase diagram' of the transition as a function of the mass ratio and the interaction strength is given. Section \hyperref[sec:quasiparticle-weight-2]{VI} provides the numerical results for the quasiparticle weight of the polaron and molecule, complementing our discussion of the polaron-to-molecule transition. The final Section \hyperref[sec:quasiparticle-weight-3]{VII} extensively discusses the numerical results for the polaron quasiparticle weight and the role of the bound state for all scattering lengths.

\section{Derivation of the Lee-Low-Pines Hamiltonian} \label{sec:LLP-transformation}

The Hamiltonian for the Fermi polaron problem, where a single impurity $\hat{d}^{\dagger}_{\mathbf{k}}$ of mass $M$ interacts with a Fermi sea of particles created by operators $\hat{c}^{\dagger}_{\mathbf{k}}$, is given by
\begin{align} \label{eq:polaron-hamiltonian-2}
    \hat{H} =\, &\sum_{\mathbf{k}} \frac{\mathbf{k}^2}{2M}\, \hat{d}^{\dagger}_{\mathbf{k}}\hat{d}_{\mathbf{k}} + \sum_{\mathbf{k}} \epsilon_\mathbf{k}\, \hat{c}^{\dagger}_{\mathbf{k}}\hat{c}_{\mathbf{k}} + \frac{g}{\mathcal{V}} \sum_{\mathbf{k}',\mathbf{k},\mathbf{q}} \hat{d}^{\dagger}_{\mathbf{k}'+\mathbf{q}}\hat{d}_{\mathbf{k}'} \hat{c}^{\dagger}_{\mathbf{k}-\mathbf{q}}\hat{c}_{\mathbf{k}} \,.
\end{align}
Here, $\epsilon_{\mathbf{k}}=\mathbf{k}^2/(2m)$ is the free dispersion relation of bath fermions with mass $m$, and we assume contact interactions with bare coupling constant $g$. In the single-impurity limit, the Hamiltonian~\eqref{eq:polaron-hamiltonian-2} can be written in terms of the impurity position and momentum operators $\hat{\mathbf{r}}$ and $\hat{\mathbf{P}}$, respectively, such that in first quantization~\cite{Lee:1953} the Hamiltonian becomes
\begin{align} \label{eq:fröhlich-hamiltonian}
    \hat{H} = \frac{\hat{\mathbf{P}}^2}{2M} + \sum_{\mathbf{k}} \epsilon_{\mathbf{k}}\, \hat{c}^{\dagger}_{\mathbf{k}}\hat{c}_{\mathbf{k}} + \frac{g}{\mathcal{V}} \sum_{\mathbf{k},\mathbf{q}} \ee^{i\mathbf{q}\cdot\hat{\mathbf{r}}}\, \hat{c}^{\dagger}_{\mathbf{k}-\mathbf{q}}\hat{c}_{\mathbf{k}} \,.
\end{align}
It is possible to decouple the impurity and bath Hilbert spaces by means of the unitary Lee-Low-Pines (LLP) transformation~\cite{Lee:1953,Kain:2017HF},
\begin{align} \label{eq:LLP-transformation}
    \hat{U} = \ee^{i\hat{\mathbf{r}}\cdot\hat{\mathbf{P}}_f} \,, \quad \hat{\mathbf{P}}_f = \sum_{\mathbf{k}} \mathbf{k}\, \hat{c}^{\dagger}_{\mathbf{k}}\hat{c}_{\mathbf{k}} \,,
\end{align}
where the total momentum operator of the fermions $\hat{\mathbf{P}}_f$ is used as generator to effectively transform to the frame comoving with the impurity. From the LLP transformation, it follows that
\begin{align} \label{eq:LLP-transformation-example}
    \hat{U}\hat{c}_{\mathbf{k}}\hat{U}^{-1} = \hat{c}_{\mathbf{k}} \ee^{-i\hat{\mathbf{r}}\cdot\mathbf{k}}\,, \quad \hat{U}\hat{\mathbf{P}}\hat{U}^{-1}=\hat{\mathbf{P}}-\hat{\mathbf{P}}_f  \,,
\end{align}
which yields the transformed Hamiltonian,
\begin{align} \label{eq:LLP-hamiltonian-1}
    \hat{\mathcal{H}}\equiv\hat{U}\hat{H}\hat{U}^{-1}= \frac{\left(\hat{\mathbf{P}}-\hat{\mathbf{P}}_f\right)^2}{2M} + \sum_{\mathbf{k}} \epsilon_{\mathbf{k}}\, \hat{c}^{\dagger}_{\mathbf{k}}\hat{c}_{\mathbf{k}} + \frac{g}{\mathcal{V}} \sum_{\mathbf{k},\mathbf{k}'} \hat{c}^{\dagger}_{\mathbf{k}}\hat{c}_{\mathbf{k}'} \,.
\end{align}
The LLP transformation makes use of the conservation of the total momentum, $\hat{\mathbf{P}}_T=\hat{\mathbf{P}}+\hat{\mathbf{P}}_f$, of the system. This becomes evident when recognizing that, in the LLP frame, the impurity momentum operator, $\hat{\mathbf{P}}=\hat{U}\hat{\mathbf{P}}_T\hat{U}^{-1}$, indeed represents the total momentum of the system. Hence, in \eqref{eq:LLP-hamiltonian-1} one can replace $\hat{\mathbf{P}}\rightarrow\mathbf{P}$, since now $[\hat{\mathbf{P}},\hat{\mathcal{H}}]=0$, i.e., $\hat{\mathbf{P}}$ has become a conserved quantity. Expanding the first term of Eq.~\eqref{eq:LLP-hamiltonian-1} and performing normal-ordering yields
\begin{align} \label{eq:HLLP}
    \hat{\mathcal{H}} = &\,\frac{\mathbf{P}^2}{2M} + \sum_{\mathbf{k}} \left(\epsilon_{\mathbf{k}} + \frac{\mathbf{k}^2}{2M} - \frac{\mathbf{k}\cdot\mathbf{P}}{M}\right)\hat{c}^{\dagger}_{\mathbf{k}}\hat{c}_{\mathbf{k}} + \frac{g}{\mathcal{V}} \sum_{\mathbf{k},\mathbf{k}'} \hat{c}^{\dagger}_{\mathbf{k}}\hat{c}_{\mathbf{k}'} + \frac{1}{2M} \sum_{\mathbf{k},\mathbf{k}'} \left(\mathbf{k}\cdot\mathbf{k}'\right) \hat{c}^{\dagger}_{\mathbf{k}}\hat{c}^{\dagger}_{\mathbf{k}'}\hat{c}_{\mathbf{k}'}\hat{c}_{\mathbf{k}} \,.
\end{align}
Due to normal-ordering, the dispersion relation of the fermions is effectively modified. For $\mathbf{P}=0$, the fermions indeed feature a modified mass given by the reduced mass $m_r=\left(1/m+1/M\right)^{-1}$ of the system. Furthermore, the quartic term (in the fermionic operators) highlights that the decoupling of the impurity and the bath comes at a cost: the originally free fermions have now become effectively interacting. In the following, we will focus on the case of zero total momentum, $\mathbf{P}=0$.

\section{Infrared divergencies and resummation} \label{sec:infrared-divergencies}

This section aims to highlight some of the theoretical challenges encountered when trying to connect the orthogonality catastrophe (OC) and physics of mobile impurities interacting with a Fermi sea.
We start by considering the ground state of the quadratic part of \cref{eq:HLLP} denoted as $|\text{GS}\rangle$. From this state, we can calculate the correction to the ground state energy at the first order of  perturbation theory with respect to the Lee-Low-Pines interaction introduced in \cref{eq:HLLP}. The correction is given by the expectation value, $\langle \text{GS}|{\frac{1}{2M}}\sum_{\mathbf{k},\mathbf{k}'}\left(\mathbf{k}\cdot\mathbf{k}'\right) \hat{c}^{\dagger}_{\mathbf{k}}\hat{c}^{\dagger}_{\mathbf{k}'}\hat{c}_{\mathbf{k}'}\hat{c}_{\mathbf{k}}|\text{GS}\rangle$. To calculate this term, we may setup the following Feynman rules,
\begin{align}
    &G_\mathbf{k}(t,t') = -i\langle \text{FS}|{\cal T}c_\mathbf{k}(t)c_\mathbf{k}^\dagger(t')|\text{FS}\rangle = -i(\theta(t-t')(1- n_F(\epsilon_\mathbf{k})) + \theta(t'-t)n_F(\epsilon_\mathbf{k}))\ee^{-i\epsilon_\mathbf{k}(t-t')}=
    \begin{tikzpicture}[baseline={([yshift=-1.8ex]current bounding box.center)}]
\begin{feynhand}
\vertex[] (a) at (-0.6,0){};
\vertex (c) at (0.6,0){};
    \graph{(a)--[fer,in=180, out=0, edge label = $\mathbf{k}$](c)};
\end{feynhand}
\end{tikzpicture},\\
&\frac{g}{\cal V} c^\dagger_\mathbf{k}c_{\mathbf{k}'} = \begin{tikzpicture}[baseline={([yshift=-.5ex]current bounding box.center)}]
\begin{feynhand}
\vertex (a) at (-0.6,0){$\mathbf{k}'$};
\vertex[dot] (b1) at (0,0){};
\vertex[crossdot] (d1) at (0,0.8){};
\vertex (c) at (0.5,0){$\mathbf{k}$};
    \graph{(a)--[fer,in=180, out=0](b1)--[fer,in=180, out=0](c), (b1)--[sca, in=270, out=90, edge label= $g$](d1)};
\end{feynhand}
\end{tikzpicture},\\
&T(\omega) = 
\begin{tikzpicture}[baseline={([yshift=-.5ex]current bounding box.center)}]
\begin{feynhand}
\vertex (a) at (-0.5,0){};
\vertex[dot] (b1) at (0,0){};
\vertex[crossdot] (d1) at (0,0.8){};
\vertex (c) at (0.5,0){};
    \graph{(a)--[fer,in=180, out=0](b1)--[fer,in=180, out=0](c), (b1)--[sca, in=270, out=90, edge label= $g$](d1)};
\end{feynhand}
\end{tikzpicture}+\begin{tikzpicture}[baseline={([yshift=-.5ex]current bounding box.center)}]
\begin{feynhand}
\vertex (a) at (-0.5,0){};
\vertex[dot] (b1) at (0,0){};
\vertex[crossdot] (d1) at (0,0.8){};
\vertex[dot] (b2) at (1,0){};
\vertex[crossdot] (d2) at (1,0.8){};
\vertex (c) at (1.5,0){};
    \graph{(a)--[fer,in=180, out=0](b1)--[fer,in=180, out=0](b2)--[fer,in=180, out=0](c), (b1)--[sca, in=270, out=90, edge label= $g$](d1),(b2)--[sca, in=270, out=90, edge label= $g$](d2)};
\end{feynhand}
\end{tikzpicture} + \begin{tikzpicture}[baseline={([yshift=-.5ex]current bounding box.center)}]
\begin{feynhand}
\vertex (a) at (-0.5,0){};
\vertex[dot] (b1) at (0,0){};
\vertex[crossdot] (d1) at (0,0.8){};
\vertex[dot] (b2) at (1,0){};
\vertex[crossdot] (d2) at (1,0.8){};
\vertex[dot] (b3) at (2,0){};
\vertex[crossdot] (d3) at (2,0.8){};
\vertex (c) at (2.5,0){};
    \graph{(a)--[fer,in=180, out=0](b1)--[fer,in=180, out=0](b2)--[fer,in=180, out=0](b3)--[fer,in=180, out=0](c), (b1)--[sca, in=270, out=90, edge label= $g$](d1),(b2)--[sca, in=270, out=90, edge label= $g$](d2),(b3)--[sca, in=270, out=90, edge label= $g$](d3)};
\end{feynhand}
\end{tikzpicture}+\cdots = \begin{tikzpicture}[baseline={([yshift=-.5ex]current bounding box.center)}]
\begin{feynhand}
\vertex[grayblob] (b) at (0,0){$T$};
\vertex (a) at (-0.8,0){};
\vertex (c) at (0.8,0){};
    \graph{(a)--[fer,in=180, out=0](b)--[fer,in=180,out=0](c)};
\end{feynhand}
\end{tikzpicture},\\
&\frac{\mathbf{k}\cdot\mathbf{k}'}{2M} \hat{c}^{\dagger}_{\mathbf{k}}\hat{c}^{\dagger}_{\mathbf{k}'}\hat{c}_{\mathbf{k}'}\hat{c}_{\mathbf{k}}=\begin{tikzpicture}[baseline=(current bounding box.center)]
\begin{feynhand}
\vertex[dot] (a) at (-0.8,0.6){};
\vertex (b1) at (-1,0){$\mathbf{k}$};
\vertex (b2) at (-1,1.2){$\mathbf{k}$};
\vertex (d1) at (1,0){$\mathbf{k}'$};
\vertex (d2) at (1,1.2){$\mathbf{k}'$};
\vertex[dot] (c) at (0.8,0.6){};
    \graph{(a)--[pho,in=180, out=0, edge label = $\frac{1}{M}\mathbf{k}\cdot\mathbf{k}'$](c), (b1)--[fer](a)--[fer](b2),(d1)--[fer](c)--[fer](d2)};
\end{feynhand}
\end{tikzpicture}.
\end{align} 
In terms of diagrams, one then has,
\begin{align}
\langle \text{GS}|{\frac{1}{2M}}\sum_{\mathbf{k},\mathbf{k}'}\left(\mathbf{k}\cdot\mathbf{k}'\right) \hat{c}^{\dagger}_{\mathbf{k}}\hat{c}^{\dagger}_{\mathbf{k}'}\hat{c}_{\mathbf{k}'}\hat{c}_{\mathbf{k}}|\text{GS}\rangle= {\frac{1}{2}}
\begin{tikzpicture}[baseline=-\the\dimexpr\fontdimen22\textfont2\relax]
\begin{feynhand}
\vertex[] (a) at (-1,0){};
\vertex[dot] (a1) at (0,0.5){};
\vertex[dot] (c1) at (0,-0.5){};
\vertex (c) at (1,0){};
    \graph{(a1)--[fer,in=0, out=0, edge label = $\mathbf{k}$](c1)--[fer,in=180,out=180, edge label = $\mathbf{k}$](a1), (a1)--[pho, in=90,out=270](c1)};
\end{feynhand}
\end{tikzpicture} + 
\begin{tikzpicture}[baseline=-\the\dimexpr\fontdimen22\textfont2\relax]
\begin{feynhand}
\vertex[grayblob] (a) at (-1,0){$T$};
\vertex[dot] (a1) at (0,0.5){};
\vertex[dot] (c1) at (0,-0.5){};
\vertex (c) at (1,0){};
    \graph{(a1)--[fer,in=0, out=0, edge label = $\mathbf{k}$](c1)--[fer,in=270,out=180, edge label = $\mathbf{k}$](a)--[fer,in=180,out=90, edge label = $\mathbf{k}$](a1), (a1)--[pho, in=90,out=270](c1)};
\end{feynhand}
\end{tikzpicture} + {\frac{1}{2}}
\begin{tikzpicture}[baseline=-\the\dimexpr\fontdimen22\textfont2\relax]
\begin{feynhand}
\vertex[grayblob] (a) at (-1,0){$T$};
\vertex[dot] (a1) at (0,0.5){};
\vertex[dot] (c1) at (0,-0.5){};
\vertex[grayblob] (c) at (1,0){$T$};
    \graph{(a1)--[fer,in=90, out=0, edge label = $\mathbf{k}$](c)--[fer,in=0, out=270, edge label = $\mathbf{p}$](c1)--[fer,in=270,out=180, edge label = $\mathbf{p}$](a)--[fer,in=180,out=90, edge label = $\mathbf{k}$](a1), (a1)--[pho, in=90,out=270](c1)};
\end{feynhand}
\end{tikzpicture}.\label{eq:1OrdPertDiag}
\end{align}
Note, in this description, the appearance of the $T$-matrix reflects the basis change to the exact single-particle eigenbasis of the quadratic Hamiltonian.  Since we consider an s-wave contact interaction, the last diagram in \cref{eq:1OrdPertDiag} vanishes due to average over the angle, $\theta_{\mathbf{k},\mathbf{p}} = \sphericalangle (\mathbf{k},\mathbf{p})$. The first diagram in \cref{eq:1OrdPertDiag} yields a constant that is independent of the interaction,
\begin{align}
    E_0 = \sum_{|\mathbf{k}|<k_F} \frac{\mathbf{k}^2}{2M} \,.
\end{align}
The second diagram in \cref{eq:1OrdPertDiag}, in turn, represents a correction to the polaron energy,
\begin{align}
    \begin{tikzpicture}[baseline=-\the\dimexpr\fontdimen22\textfont2\relax]
\begin{feynhand}
\vertex[grayblob] (a) at (-1,0){$T$};
\vertex[dot] (a1) at (0,0.5){};
\vertex[dot] (c1) at (0,-0.5){};
\vertex (c) at (1,0){};
    \graph{(a1)--[fer,in=0, out=0, edge label = $\mathbf{k}$](c1)--[fer,in=270,out=180, edge label = $\mathbf{k}$](a)--[fer,in=180,out=90, edge label = $\mathbf{k}$](a1), (a1)--[pho, in=90,out=270](c1)};
\end{feynhand}
\end{tikzpicture} = \sum_{|\mathbf{k}|<k_F} \mathbf{k}^2 \int_{-\infty}^\infty d\lambda ~{\frac{1-n_{F}(\lambda)+(\lambda-\epsilon_{\mathbf{k}})\partial_{\epsilon_{\mathbf{k}}} n_{F}(\epsilon_{\mathbf{k}})}{(\lambda- \epsilon_{\mathbf{k}})^2}}{\frac{1}{\pi}}\text{Im}{T}(\lambda) \,.
\end{align}
Crucially, after carrying out the integration over $\lambda$, the remaining sum over momenta $|\mathbf{k}|<k_F$, in the form of $\sum_{|\mathbf{k}|<k_F} \left(A+\frac{B}{E_F-\epsilon_{\bf k}}\right)$, is logarithmically divergent in the limit $\epsilon_{\mathbf{k}} \rightarrow E_F$.

In order to cure these logarithmic divergencies, one may rather consider an effective Hamiltonian where contributions from the quartic LLP term are already partially resummed leading to a gap that can remove the IR divergencies. To find such a description, one recognizes that the first-order correction to the fermion self-energy is given by
\begin{align}
\begin{tikzpicture}[baseline=-\the\dimexpr\fontdimen22\textfont2\relax]
\begin{feynhand}
\vertex (a) at (0,0){};
\vertex[dot] (a1) at (0.5,0){};
\vertex[dot] (c1) at (1.5,0){};
\vertex (c) at (2,0){};
    \graph{(a)--[fer,in=180, out=0, edge label' = $\mathbf{k}$](a1)--[fer,in=180,out=0, edge label' = $\mathbf{k}$](c1)--[fer,in=180,out=0, edge label' = $\mathbf{k}$](c), (a1)--[pho, in=90,out=90, edge label=$\frac{1}{M}\mathbf{k}\cdot\mathbf{k}$](c1)};
\end{feynhand}
\end{tikzpicture} =-\frac{1}{M}
\mathbf{k}^2n_F(\epsilon_{\mathbf{k}})
\,.\label{eq:slf}
\end{align}

Importantly, in a particle-hole picture (where holes denote the unoccupied states below the Fermi sea), the above diagram, due to the appearance of the Fermi distribution function, generates a change in the dispersion relation of the holes. This change is different for the particle states above the Fermi surface,  leading to a modified dispersion relation which has the same expression as given by \cref{eq:gappedDisp} below (note, that the self-energy correction applies to the already modified dispersion relation of fermions in Eq.~\eqref{eq:HLLP}). 
Noticing that the self-energy diagram does not involve interactions between the impurity and the bath fermions, the different dispersion energies for particle-like and hole-like excitations are a fundamental property of momentum conservation, regardless of the interaction potential. The role of self-energy \cref{eq:slf} in modifying the particle and hole dispersion relations becomes also directly evident when considering how the kinetic energy operator in \cref{eq:LLP-hamiltonian-1} (at total momentum ${\mathbf{P}}=0$) acts on a single particle (hole) excitation of the unperturbed Fermi sea. 
\begin{align}
    &\left[\frac{\hat{\mathbf{P}}_f^2}{2M} + \sum_{\mathbf{k'}} \epsilon_{\mathbf{k'}}\, \hat{c}^{\dagger}_{\mathbf{k'}}\hat{c}_{\mathbf{k'}}  \right] \hat{c}_{\mathbf{k}}^\dagger|\text{FS}\rangle = \left(\frac{{\mathbf{k}}^2}{2M} + \frac{{\mathbf{k}}^2}{2m}\right)\hat{c}_{\mathbf{k}}^\dagger|\text{FS}\rangle +E_{\text{FS}}\hat{c}_{\mathbf{k}}^\dagger|\text{FS}\rangle \,, \label{eq:1pExEng}\\
    &\left[\frac{\hat{\mathbf{P}}_f^2}{2M} + \sum_{\mathbf{k'}} \epsilon_{\mathbf{k'}}\, \hat{c}^{\dagger}_{\mathbf{k'}}\hat{c}_{\mathbf{k'}}  \right] \hat{c}_{\mathbf{q}}|\text{FS}\rangle = \left(\frac{{\mathbf{q}}^2}{2M} - \frac{{\mathbf{q}}^2}{2m}\right)\hat{c}_{\mathbf{q}}|\text{FS}\rangle +E_{\text{FS}}\hat{c}_{\mathbf{q}}|\text{FS}\rangle \,\label{eq:1hExEng},
\end{align}
where $E_{\text{FS}}=\sum_{|\mathbf{k}|<k_F} \epsilon_\mathbf{k}$, and we used the fact that $\hat{c}_\mathbf{k}|\text{FS}\rangle$ is an eigenstate of the operator $\hat{\mathbf{P}}_f$. Inspecting \cref{eq:1pExEng} and \cref{eq:1hExEng}, the role of self-energy \cref{eq:slf} can be trivially identified. Since \cref{eq:1pExEng} and \cref{eq:1hExEng} are exact, the one loop self-energy \cref{eq:slf} is exact, and one has
\begin{align}
    E_{\mathbf{k}} =
    \begin{cases}
        \frac{\mathbf{k}^2}{2m} - \frac{\mathbf{k}^2}{2M} & \quad\mathrm{for}\quad |\mathbf{k}|<k_F \,, \\[1ex]
        \frac{\mathbf{k}^2}{2m} + \frac{\mathbf{k}^2}{2M} & \quad\mathrm{for}\quad |\mathbf{k}|>k_F \,.
    \end{cases}\label{eq:gappedDisp}
\end{align}
Note that we find this property already in the non-interacting problem. This motivates us to reorder the Hamiltonian in such a way to account for this self-energy effect already on an operator level, as explicitly done in the following section.

\section{Reordering and gapped dispersion} \label{sec:gapped-dispersion}

Following the idea on how to cure the infrared divergence by incooperating the self-energy effect as described in \cref{eq:slf}, we split the momentum sums in the quartic LLP  term of the Hamiltonian \cref{eq:HLLP} into hole contributions with $|\mathbf{q}|<k_F$ and particle contributions with $|\mathbf{k}|>k_F$, i.e., rewriting the momentum sums as $\sum_{\mathbf{k}} = \sum_{|\mathbf{q}|<k_F} + \sum_{|\mathbf{k}|>k_F}$. Grouping the corresponding terms, results in the expression
\begin{align}\nonumber \label{eq:LLP-hamiltonian-ph-split}
    \hat{H} = &\sum_{|\mathbf{q}|<k_F} \left(\frac{\mathbf{q}^2}{2m}+\frac{\mathbf{q}^2}{2M}\right)\hat{c}^{\dagger}_{\mathbf{q}}\hat{c}_{\mathbf{q}} + \sum_{|\mathbf{k}|>k_F} \left(\frac{\mathbf{k}^2}{2m}+\frac{\mathbf{k}^2}{2M}\right)\hat{c}^{\dagger}_{\mathbf{k}}\hat{c}_{\mathbf{k}} \\
    &+ \frac{1}{2M} \sum_{|\mathbf{q}|,|\mathbf{q}'|<k_F} \left(\mathbf{q}\cdot\mathbf{q}'\right) \hat{c}^{\dagger}_{\mathbf{q}}\hat{c}^{\dagger}_{\mathbf{q}'}\hat{c}_{\mathbf{q}'}\hat{c}_{\mathbf{q}} + \frac{1}{M} \sum_{|\mathbf{k}|>k_F,|\mathbf{q}|<k_F} \left(\mathbf{k}\cdot\mathbf{q}\right) \hat{c}^{\dagger}_{\mathbf{k}}\hat{c}^{\dagger}_{\mathbf{q}}\hat{c}_{\mathbf{q}}\hat{c}_{\mathbf{k}} + \frac{1}{2M} \sum_{|\mathbf{k}|,|\mathbf{k}'|>k_F} \left(\mathbf{k}\cdot\mathbf{k}'\right) \hat{c}^{\dagger}_{\mathbf{k}}\hat{c}^{\dagger}_{\mathbf{k}'}\hat{c}_{\mathbf{k}'}\hat{c}_{\mathbf{k}} \,.
\end{align}
Next, we reorder the three resulting quartic LLP terms in such a way that their quartic contributions evaluate to zero when acting on the free Fermi sea $\ket{\text{FS}}$. For instance, for the first term, this yields
\begin{align}
    \sum_{|\mathbf{q}|,|\mathbf{q}'|<k_F} \frac{\left(\mathbf{q}\cdot\mathbf{q}'\right) }{2M}\hat{c}^{\dagger}_{\mathbf{q}}\hat{c}^{\dagger}_{\mathbf{q}'}\hat{c}_{\mathbf{q}'}\hat{c}_{\mathbf{q}}=\sum_{|\mathbf{q}|<k_F} \frac{\mathbf{q}^2}{2M}-2\sum_{|\mathbf{q}|<k_F} \frac{\mathbf{q}^2}{2M}\hat{c}^\dagger_{\mathbf{q}}\hat{c}_{\mathbf{q}}+\sum_{|\mathbf{q}|,|\mathbf{q}'|<k_F} \frac{\left(\mathbf{q}\cdot\mathbf{q}'\right) }{2M}\hat{c}_{\mathbf{q}}\hat{c}_{\mathbf{q}'}\hat{c}^{\dagger}_{\mathbf{q}'}\hat{c}^{\dagger}_{\mathbf{q}} \,,
\end{align}
where $(1/2M)\sum_{|\mathbf{q}|,|\mathbf{q}'|<k_F} \left(\mathbf{q}\cdot\mathbf{q}'\right)\hat{c}_{\mathbf{q}}\hat{c}_{\mathbf{q}'}\hat{c}^{\dagger}_{\mathbf{q}'}\hat{c}^{\dagger}_{\mathbf{q}}\ket{\text{FS}}=0$. Following the same prescription for the other two terms, we obtain the Hamiltonian
\begin{align} \label{eq:effective-Hamiltonian-reordered}
    \hat{H} =\, &\,\sum_{|\mathbf{q}|<k_F} \frac{\mathbf{q}^2}{2M} + \sum_{|\mathbf{q}|<k_F} \left(\frac{\mathbf{q}^2}{2m}-\frac{\mathbf{q}^2}{2M}\right)\hat{c}^{\dagger}_{\mathbf{q}}\hat{c}_{\mathbf{q}} + \sum_{|\mathbf{k}|>k_F} \left(\frac{\mathbf{k}^2}{2m}+\frac{\mathbf{k}^2}{2M}\right)\hat{c}^{\dagger}_{\mathbf{k}}\hat{c}_{\mathbf{k}} + \frac{g}{\mathcal{V}} \sum_{\mathbf{k},\mathbf{k}'} \hat{c}^{\dagger}_{\mathbf{k}}\hat{c}_{\mathbf{k}'} \notag \\
    &+ \frac{1}{2M} \sum_{|\mathbf{q}|,|\mathbf{q}'|<k_F} \left(\mathbf{q}\cdot\mathbf{q}'\right) \hat{c}_{\mathbf{q}}\hat{c}_{\mathbf{q}'} \hat{c}^{\dagger}_{\mathbf{q}'}\hat{c}^{\dagger}_{\mathbf{q}} - \frac{1}{M} \sum_{|\mathbf{k}|>k_F,|\mathbf{q}|<k_F} \left(\mathbf{k}\cdot\mathbf{q}\right) \hat{c}^{\dagger}_{\mathbf{k}}\hat{c}_{\mathbf{q}}\hat{c}^{\dagger}_{\mathbf{q}}\hat{c}_{\mathbf{k}} + \frac{1}{2M} \sum_{|\mathbf{k}|,|\mathbf{k}'|>k_F} \left(\mathbf{k}\cdot\mathbf{k}'\right) \hat{c}^{\dagger}_{\mathbf{k}}\hat{c}^{\dagger}_{\mathbf{k}'}\hat{c}_{\mathbf{k}'}\hat{c}_{\mathbf{k}} \,,
\end{align}
which defines the mass-gap model of \cref{eq:gapped-Hamiltonian} in the main text. Note that we find again the gapped dispersion $E_{\mathbf{k}}$, which arises from the fermionic anti-commutation relation. This cures the infrared divergencies in the perturbative approaches, enabling us to now connect the orthogonality catastrophe with the physics of mobile Fermi polarons in one unified framework.

\section{Exact solution of the quadratic Hamiltonian} \label{sec:exact-solution}

We can find an approximation of the ground state energy by diagonalizing the quadratic part of the Hamiltonian $\hat{\mathcal{H}}$, given by
\begin{align} \label{eq:quadratic-hamiltonian}
    \hat{\mathcal{H}}_{\mathrm{quad}} = \sum_{\mathbf{k}} E_{\mathbf{k}} \hat{c}^{\dagger}_{\mathbf{k}}\hat{c}_{\mathbf{k}} + \frac{g}{\mathcal{V}} \sum_{\mathbf{k},\mathbf{k}'} \hat{c}^{\dagger}_{\mathbf{k}}\hat{c}_{\mathbf{k}'} \,,
\end{align}
and spanning a Fermi sea $|\widetilde{\text{FS}}\rangle$ from the corresponding single-particle eigenstates. At the end of the section, we show that this procedure yields a strict upper variational bound to the true ground state energy of $\hat{\mathcal{H}}$ (and thus $\hat H$) because the quartic terms in \cref{eq:effective-Hamiltonian-reordered} evaluate to zero when acting on the interacting Fermi sea $|\widetilde{\text{FS}}\rangle$.

The energy spectrum of $\hat{\mathcal{H}}_{\mathrm{quad}}$ can be computed either numerically via exact diagonalization or analytically using the scattering $T$-matrix, as demonstrated below. We write the diagonalized Hamiltonian as
\begin{align} \label{eq:diagonal-quadratic-hamiltonian}
    \hat{\mathcal{H}}_{\mathrm{quad}} = \sum_{\bm{\nu}} \omega_{\bm{\nu}} \hat{\gamma}^{\dagger}_{\bm{\nu}} \hat{\gamma}_{\bm{\nu}} \,,
\end{align}
where $\bm{\nu}$ labels the (possibly continuous) diagonalized eigenstates with energies $\omega_{\bm{\nu}}$. The new operators $\hat{\gamma}^{(\dagger)}_{\bm{\nu}}$ are connected to the original non-interacting fermions $\hat{c}^{(\dagger)}_{\mathbf{k}}$ via a unitary basis change transformation,
\begin{align} \label{eq:basis-change-operators}
    \hat{\gamma}^{\dagger}_{\bm{\nu}} = \sum_{\mathbf{k}} \langle\mathbf{k}|\tilde{\bm{\nu}}\rangle \hat{c}^{\dagger}_{\mathbf{k}} \,,
\end{align}
where, as mentioned before, $|\mathbf{k}\rangle=\hat{c}_{\mathbf{k}}^\dagger|0\rangle$ ($|\tilde{\bm{\nu}}\rangle=\hat{\gamma}_{\bm{\nu}}^\dagger|0\rangle$) label the single-particle orbitals of the fermions in the absence (presence) of the impurity-fermion potential.

As described in the main text, the new eigenstates $|\tilde{\bm{\nu}}\rangle$ comprise a bound state \circleletter{b}, an in-gap state \circleletter{g}, and scattering states below and above the Fermi energy. Here we intentionally distinguished the states $|\tilde{\bm{\nu}}\rangle$ from $|\nu\rangle$ used in the main text, as $|\nu\rangle$ formally label eigenstates obtained from the numerical diagonalization of the Hamiltonian in a finite-size system, while $|\tilde{\bm{\nu}}\rangle$ can also label continuous diagonalized eigenstates. A numerical solution of the energy spectrum as function of interaction $1/(k_Fa)$ is shown in Fig.~2 of the main text.  

The energy of the bound and in-gap state can be identified as the poles of the underlying $T$-matrix. For the Hamiltonian~\eqref{eq:quadratic-hamiltonian} with gapped dispersion relation $E_{\mathbf{k}}$ and contact interaction $g$, the $T$-matrix is obtained from the Lippmann-Schwinger equation, 
\begin{align} \label{eq:new-T-matrix}
    \tilde{T}(E) = \frac{1/\mathcal{V}}{\frac{1}{g} - \frac{1}{\mathcal{V}}\sum_{\mathbf{p}}\frac{1}{E - E_{\mathbf{p}} + i0^+}} = \frac{2\pi^2}{\mathcal{V}} &\left[ m_{p}\sqrt{2m_{p}E}\log\left( \frac{\sqrt{2m_{p}E}+k_F}{\sqrt{2m_{p}E}-k_F} \right) \right. \notag  \\
    &- m_{h}\sqrt{2m_{h}E}\log\left( \frac{\sqrt{2m_{h}E}+k_F}{\sqrt{2m_{h}E}-k_F} \right) \notag \\
    &\left. +\, \frac{4m_{p}m_{h}k_F}{M} + \frac{\pi m_{p}}{a_s} + i\pi m_{p} \sqrt{2m_{p}E} \right]^{-1} \,,
\end{align}
where $m_{p}=mM/(M+m)$ and $m_{h}=mM/(M-m)$ are the particle and hole mass, respectively. From this, the polaron energy can be computed exactly from the sum over all energy level shifts leading to Fumi's theorem~\cite{Fumi:1955}, 
\begin{align} \label{eq:Fumis-theorem}
    \Delta E = \sum_{\bm{\alpha}<} \omega_{\bm{\alpha}} - \sum_{|\mathbf{q}|<k_F} E_{\mathbf{q}} = E_b-\int_0^{E_F} dE\, \frac{\tilde{\delta}(E)}{\pi} \,.
\end{align}
Here $\bm{\alpha}<$ denotes the occupied single-particle states and $E_b<0$ is the bound state energy at finite impurity mass. In turn, the scattering phase shift $\tilde{\delta}(E)$ is connected to the $T$-matrix via 
\begin{align} \label{eq:phase-shift}
    \tilde{\delta}(E) = \mathrm{Im}\log \tilde{T}(E) \,.
\end{align}
As shown in \cref{fig:spectrum-plot}, being calculated for particles with a modified dispersion relation, the form of $\tilde{\delta}(E)$ differs strongly from its vacuum form $\cot(\delta(E))=-1/(a_s\sqrt{2m_rE})$. We note that for the infinite mass limit, $M\to \infty$, $\tilde{\delta}(E)$ becomes identical to the vacuum form $\delta(E)$. Most importantly, we see that the phase shift goes to zero as the scattering energy approaches the mass gap.

For completeness, we note that the overlap $\langle \mathbf{k}|\tilde{\bm \nu}\rangle$ appearing in \cref{eq:basis-change-operators} is  related to the $T$-matrix by $(E_{\bm{\nu}} - E_{\bf{k}})\langle \mathbf{k}|\tilde{\bm{\nu}}\rangle = \langle \mathbf{k}|\tilde{T}(E_{\bm{\nu}})|{\bm{\nu}}\rangle_0$, where $|{\bm{\nu}}\rangle_0=\hat{c}_{\bm{\nu}}^\dagger|0\rangle$ represents the plane wave state with momentum $\bm{\nu}$. It should be pointed out that $|\mathbf{k}\rangle$ has an identical definition with $|{\bm\nu}\rangle_0$ except that $\bm \nu$ denotes the momentum vector of the scattering wavefunctions, while $\mathbf{k}$ denotes the momentum vector for the non-interacting wavefunctions.

\begin{figure}[th]
    \centering
    \includegraphics[width=0.6\textwidth]{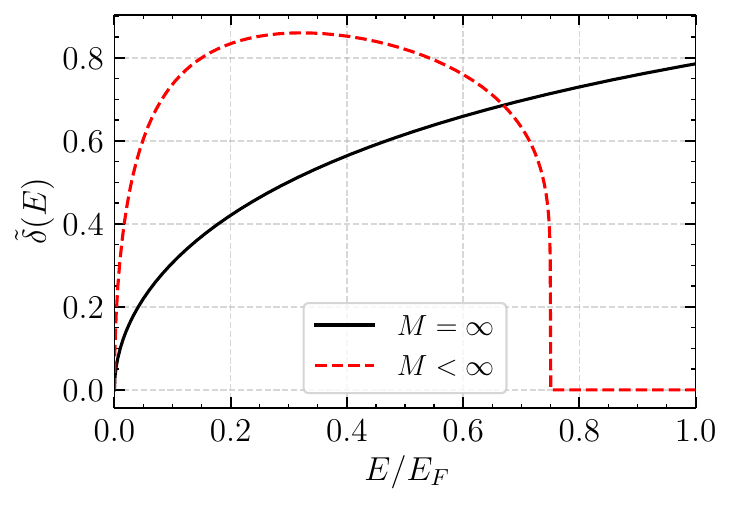}
    \caption{Modified scattering phase shift $\tilde{\delta}(E)$ for finite-mass particles with gapped dispersion relation $E_{\mathbf{k}}$ obtained from Eq.~\eqref{eq:new-T-matrix} and~\eqref{eq:phase-shift}.}
    \label{fig:spectrum-plot}
\end{figure}

We now show that the ground state of $\hat{\mathcal{H}}_{\mathrm{quad}}$ satisfies $\langle \widetilde{\text{FS}}|\hat{\cal H}_\text{int}^\text{LLP}|\widetilde{\text{FS}}\rangle=0$. This is crucial, as $\langle\widetilde{\text{FS}}|{\cal H}^\text{LLP}_\text{int}|\widetilde{\text{FS}}\rangle=0$ implies that $\tilde{E}_\text{GS}=\langle \widetilde{\text{FS}}|{ \cal H}_\text{quad}|\widetilde{\text{FS}}\rangle$ provides an upper variational bound to the true ground state energy, for the case of s-wave contact interactions. To show $\langle \widetilde{\text{FS}}|\hat{\cal H}_\text{int}^\text{LLP}|\widetilde{\text{FS}}\rangle=0$, we consider each contribution arising from \cref{eq:effective-Hamiltonian-reordered} separately, the first quartic term reads

\begin{align} \label{eq:expectation-quartic-terms}
    \sum_{|\mathbf{q}|,|\mathbf{q}'|<k_F} \left(\mathbf{q}\cdot\mathbf{q}'\right) \langle\widetilde{\text{FS}}|\hat{c}_{\mathbf{q}}\hat{c}_{\mathbf{q}'} \hat{c}^{\dagger}_{\mathbf{q}'}\hat{c}^{\dagger}_{\mathbf{q}} |\widetilde{\text{FS}}\rangle &= \sum_{|\mathbf{q}|,|\mathbf{q}'|<k_F} \left(\mathbf{q}\cdot\mathbf{q}'\right) \sum_{\bm{\alpha}\bm{\beta}\bm{\gamma}\bm{\delta}} \langle\tilde{\bm{\alpha}}|\mathbf{q}\rangle \langle\tilde{\bm{\beta}}|\mathbf{q'}\rangle \langle\mathbf{q'}|\tilde{\bm{\gamma}}\rangle \langle\mathbf{q}|\tilde{\bm{\delta}}\rangle \langle\widetilde{\text{FS}}|\hat{\gamma}_{\bm{\alpha}}\hat{\gamma}_{\bm{\beta}} \hat{\gamma}^{\dagger}_{\bm{\gamma}}\hat{\gamma}^{\dagger}_{\bm{\delta}} |\widetilde{\text{FS}}\rangle \notag \\
    &= \sum_{|\mathbf{q}|,|\mathbf{q}'|<k_F} \left(\mathbf{q}\cdot\mathbf{q}'\right) \sum_{\bm{\alpha}\bm{\beta}\bm{\gamma}\bm{\delta}} \langle\tilde{\bm{\alpha}}|\mathbf{q}\rangle \langle\tilde{\bm{\beta}}|\mathbf{q'}\rangle \langle\mathbf{q'}|\tilde{\bm{\gamma}}\rangle \langle\mathbf{q}|\tilde{\bm{\delta}}\rangle \left( \delta_{\bm{\alpha\delta},>} \delta_{\bm{\beta\gamma},>} - \delta_{\bm{\alpha\gamma},>} \delta_{\bm{\beta\delta},>} \right) \, . 
\end{align}
Here the $<$ ($>$) makes explicit that the corresponding states must be above (below) the Fermi surface. Executing the Kronecker delta functions yields
\begin{align}\label{eq:expectation-quartic-terms-2}
    \text{\cref{eq:expectation-quartic-terms}}
    &= -\sum_{|\mathbf{q}|,|\mathbf{q}'|<k_F} \left(\mathbf{q}\cdot\mathbf{q}'\right) \sum_{\bm{\alpha},\bm{\beta}>} \langle\tilde{\bm{\alpha}}|\mathbf{q}\rangle \langle\tilde{\bm{\beta}}|\mathbf{q'}\rangle \langle\mathbf{q'}|\tilde{\bm{\alpha}}\rangle \langle\mathbf{q}|\tilde{\bm{\beta}}\rangle \notag \\
    &=  -\sum_{{\bm \alpha},{\bm \beta}>} \quad \sum_{|\mathbf{q}|,|\mathbf{q}'|<k_F} \left(\mathbf{q}\cdot\mathbf{q}'\right)f(|\mathbf{q}'|,|{\bm \alpha}|)f^*(|\mathbf{q}|,|{\bm \alpha}|)f(|\mathbf{q}|,|{\bm \beta}|)f^*(|\mathbf{q}'|,|{\bm \beta}|) \, ,
\end{align}
where we defined the overlaps $\langle\mathbf{k}|\tilde{\bm{\alpha}}\rangle=f(|\mathbf{k}|,|{\bm{\alpha}}|)$. For contact interactions, these do not depend on the angles between the vectors $\mathbf{k}$ and ${\bm{\alpha}}$, in the case of $|\mathbf{k}|\neq|{\bm \alpha}|$. Hence, performing the angular integration directly proves $\text{\cref{eq:expectation-quartic-terms-2}}=\sum_{|\mathbf{q}|,|\mathbf{q}'|<k_F} \left(\mathbf{q}\cdot\mathbf{q}'\right)f(|\mathbf{q}'|,|{\bm \alpha}|)f^*(|\mathbf{q}|,|{\bm \alpha}|)f(|\mathbf{q}|,|{\bm \beta}|)f^*(|\mathbf{q}'|,|{\bm \beta}|)=0$. The evaluation of the other quartic terms in $\hat{\cal H}_\text{int}^\text{LLP}$ follows analogous steps, showing that $\langle\widetilde{\text{FS}}|\hat{\cal H}_\text{int}^{\text{LLP}}|\widetilde{\text{FS}}\rangle=0$. This, in turn,  validates  
\begin{align}
\langle\widetilde{\text{FS}}|\hat{\cal H}|\widetilde{\text{FS}}\rangle=\langle\widetilde{\text{FS}}|\hat{\cal H}_\text{quad}|\widetilde{\text{FS}}\rangle+\langle\widetilde{\text{FS}}|\hat{\cal H}_\text{int}^\text{LLP}|\widetilde{\text{FS}}\rangle+E_0=\langle\widetilde{\text{FS}}|\hat{\cal H}_\text{quad}|\widetilde{\text{FS}}\rangle+E_0 \,.
\end{align}
Recognizing that $|\widetilde{\text{FS}}\rangle$ can be regarded as a variational trial state, $\langle \widetilde{\text{FS}}|\hat{\cal H}|\widetilde{\text{FS}}\rangle$ is above the true ground state energy; in other words, it sets an upper bound of the true ground state energy of $\hat{\cal H}$, and, since $U$ is unitary, also $\hat H$.

\section{In-gap state and polaron-to-molecule transition} \label{sec:polaron-to-molecule-2}

Here we provide the details on the calculation of the molecule energy through the occupation of the in-gap state in the mass-gap description. In general, the polaron and molecule energies are given by the sum of all energy level shifts due to interactions.

\begin{figure}[b!]
    \centering
    \includegraphics[width=0.7\textwidth]{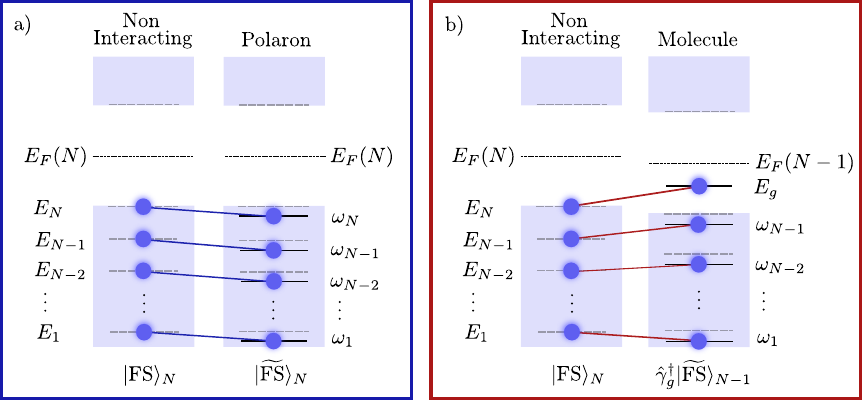}
    \caption{Construction of the ground state wave functions and energy level shifts for the attractive polaron and the molecule. a) The attractive polaron is given by occupying the lowest $N$ states of the diagonalized gapped Hamiltonian, i.e., $|\text{pol}\rangle = |\widetilde{\text{FS}}\rangle_N$. Note that $\omega_1$ may also correspond to the bound state below the lower band. b) The molecule is obtained by occupying all $N-1$ particles of a Fermi sea $|\widetilde{\text{FS}}\rangle_{N-1}$ and additionally occupying the in-gap state,  i.e., $|\text{mol}\rangle = \hat \gamma_g^\dagger |\widetilde{\text{FS}}\rangle_{N-1}$. In both calculations, the energy level shifts are computed, however,  with respect to the non-interacting (gapped) Fermi sea of $N$ particles. This leads to an effective reduction of the molecular energy by $E_F$, see \cref{eq:mol-pol-ef}.}
    \label{fig:polaron-molecule-energies}
\end{figure}

Let us first focus on the polaron energy shift with respect to a Fermi sea of $N$ particles. Using the quadratic Hamiltonian, it is given by
\begin{align}
    E_{\text{pol}} &= \left[E_0(N) + \sum_{n=1}^N \omega_n\right] - \left[E_0(N) + \sum_{n=1}^N E_n\right] \notag \\
    &= \sum_{n=1}^N (\omega_n-E_n) \equiv \Delta E(N) \,,
\end{align}
where $E_0(N)=\sum_{|\mathbf{q}|<k_F(N)} \mathbf{q}^2/(2M)$, $\omega_n$ are the interacting eigenvalues and $E_n$ are the non-interacting eigenvalues of the single-particle Hamiltonian with the gapped dispersion corresponding to $\hat{\mathcal{H}}_\text{quad}$. Thus, the polaron energy is the sum of all energy level shifts due to interactions, see Fig.~\ref{fig:polaron-molecule-energies}(a).

In order to construct the molecule state we  start with $N-1$ particles in the Fermi sea and additionally occupy the in-gap state as shown in Fig.~\ref{fig:polaron-molecule-energies}(b). Then the molecule energy is given by
\begin{align}
    E_{\text{mol}} &= \left[E_0(N-1) + \sum_{n=1}^{N-1} \omega_n + E_g\right] - \left[E_0(N) + \sum_{n=1}^N E_n\right] \notag \\
    &= E_0(N-1)-E_0(N) + \Delta E(N-1) + E_g - E_N \,.
\end{align}
Note that we still compare the total energy to the non-interacting Fermi sea with $N$ particles. In the continuum limit, we have $E_0(N-1)=E_0(N)-k^2_F/(2M)$, $\Delta E(N-1) = \Delta E(N)$ and $E_N=E_F-k^2_F/(2M)$. Thus, we obtain
\begin{align}
    E_{\text{mol}} = E_{\text{pol}} + E_g - E_F \,. \label{eq:mol-pol-ef}
\end{align}
From this we see that the molecule is the ground state for $E_g<E_F$. We show the polaron and molecule energy computed for different mass ratios and interaction strengths in Fig.~\ref{fig:polaron-molecule-phase-diagram}.

\begin{figure}[h!]
    \includegraphics[width=0.75\textwidth]{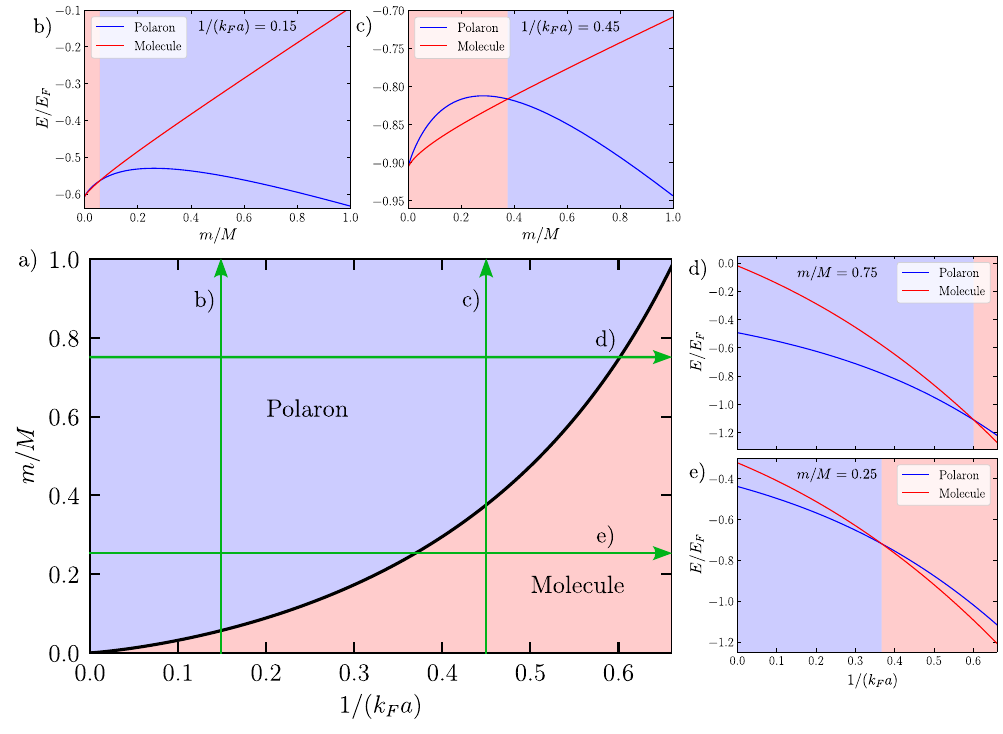}
    \caption{a) Polaron-to-molecule phase diagram as defined by the polaron and molecule energy (cuts shown in subfigures b-e) as a function of inverse mass ratio $m/M$ and interaction strength $1/(k_Fa)$.}
    \label{fig:polaron-molecule-phase-diagram}
\end{figure}

\section{Quasiparticle weight and overlap between polaron and molecule} \label{sec:quasiparticle-weight-2}

\begin{figure}[t!]
    \centering
    \includegraphics[width=0.84\textwidth]{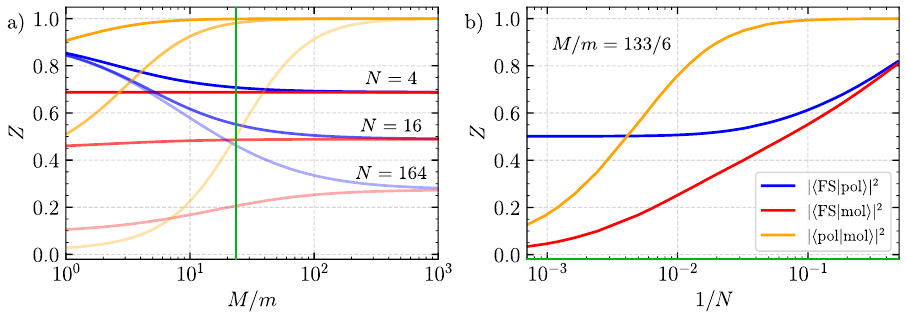}
    \caption{Analysis of the quasiparticle weight $Z_{\text{pol}}=|\langle\mathrm{FS}|\text{pol}\rangle|^2$ (blue line), $Z_{\text{mol}}=|\langle\mathrm{FS}|\text{mol}\rangle|^2$ (red line) and the overlap $|\langle\text{pol}|\text{mol}\rangle|^2$ (orange line) as function of fermion number $N$ and mass ratio $M/m$. a) Numerical results for constant interaction strength $1/(k_Fa)=0$ as function of mass ratio $M/m$ for different system sizes $N$. b) Overlaps for constant interaction strength $1/(k_Fa)=0$ and fixed mass ratio $M/m=133/6$ of a Cs-Li mixture as function of $N$.}
    \label{fig:polaron-molecule-overlap}
\end{figure}
Within the mass-gap description, we can explicitly define the polaron and molecule wave function, $|\text{pol}\rangle = |\widetilde{\text{FS}}\rangle_N$ and
$|\text{mol}\rangle = \hat \gamma_g^\dagger |\widetilde{\text{FS}}\rangle_{N-1}$,
respectively. Using the procedure described above, we compute the eigenenergies $\omega_\nu$ and eigenvectors $\langle n\vert \nu\rangle$ for a finite basis set via exact diagonalization for the polaron and molecule. The quasiparticle weight is then given by the Slater determinant $Z_{\text{pol}}=|\langle\mathrm{FS}|\text{pol}\rangle|^2$ and $Z_{\text{mol}}=|\langle\mathrm{FS}|\text{mol}\rangle|^2$, as also described in the main text. The overlap $|\langle\text{pol}|\text{mol}\rangle|^2$ is, in turn, obtained by
\begin{align}
    |\langle\text{pol}|\text{mol}\rangle|^2 = \det
    \begin{bmatrix}
        \langle \nu_{\text{pol}}=1|\nu_{\text{mol}}=1 \rangle & \cdots & \langle \nu_{\text{pol}}=1|\nu_{\text{mol}}=N \rangle \\
        \vdots &  \ddots & \vdots  \\
        \langle \nu_{\text{pol}}=N|\nu_{\text{mol}}=1 \rangle & \cdots & \langle \nu_{\text{pol}}= N|\nu_{\text{mol}} =N\rangle
    \end{bmatrix}^2  \,,
\end{align}
where $N$ is the highest occupied state, and $\nu_{\text{pol}}$ and $\nu_{\text{mol}}$ label the quantum numbers of the single-particle orbitals of the polaron and molecule wave function, respectively. Fig.~\ref{fig:polaron-molecule-overlap} shows the overlaps $|\langle\mathrm{FS}|\text{pol}\rangle|^2$, $|\langle\mathrm{FS}|\text{mol}\rangle|^2$ and $|\langle\text{pol}|\text{mol}\rangle|^2$ for different system sizes $N$ as a function of the interaction strength $1/(k_Fa)$ and the mass ratio $M/m$.

In Fig.~\ref{fig:polaron-molecule-overlap}a), we observe that the overlap $|\langle\text{pol}|\text{mol}\rangle|^2 \rightarrow 1$ as the mass ratio $M/m \rightarrow \infty$. For finite system sizes where $N<\infty$, there exists a finite mass ratio $M/m<\infty$ for which $|\langle\text{pol}|\text{mol}\rangle|^2=1$. This can be understood within the mass-gap description as follows: if the mass ratio becomes too large (i.e., the mass gap $\Delta(M)$ becomes small), the system can no longer resolve the mass gap. As a result, the in-gap state vanishes, and with it the distinction between polaron and molecule state --- rendering them effectively degenerate. Consequently, there exists a characteristic ratio $Nm/M$, dependent on $1/(k_Fa)$, for which the polaron and molecule states become indistinguishable.

In Fig.~\ref{fig:polaron-molecule-overlap}b), we show the dependence of $Z$ as a function of $1/N$, one clearly sees that the molecular $Z$ factor vanishes in the continuum limit, while the polaron retains a finite quasiparticle weight for any fixed interaction strength and mass ratio. As the system size $N$ is decreased, the polaron and molecule state become indistinguishable. For completeness, in Fig.~\ref{fig:Z-factor-ground-state}, we show the quasiparticle weight $Z$ of the ground state as function of $1/(k_Fa)$ in the continuum limit for the mass-balanced case, as discussed in the main text.

\begin{figure}[h!]
    \centering
    \includegraphics[width=0.82\textwidth]{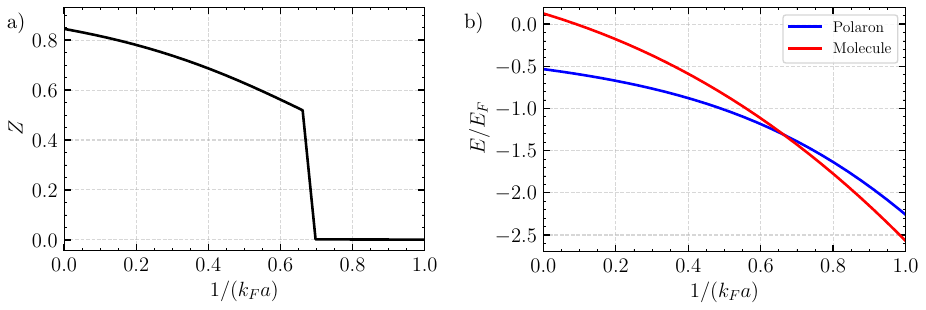}
    \caption{a) Quasiparticle weight $Z$ of the ground state and b) polaron (molecule) energy as function of $1/(k_Fa)$  at mass balance. One clearly sees that the $Z$ factor drops to zero once the molecule becomes the ground state of the system.}
    \label{fig:Z-factor-ground-state}
\end{figure}

\section{Power-law decay of the  quasiparticle weight as function of mass} \label{sec:quasiparticle-weight-3}

\begin{figure}[t!]
    \centering
    \includegraphics[width=\textwidth]{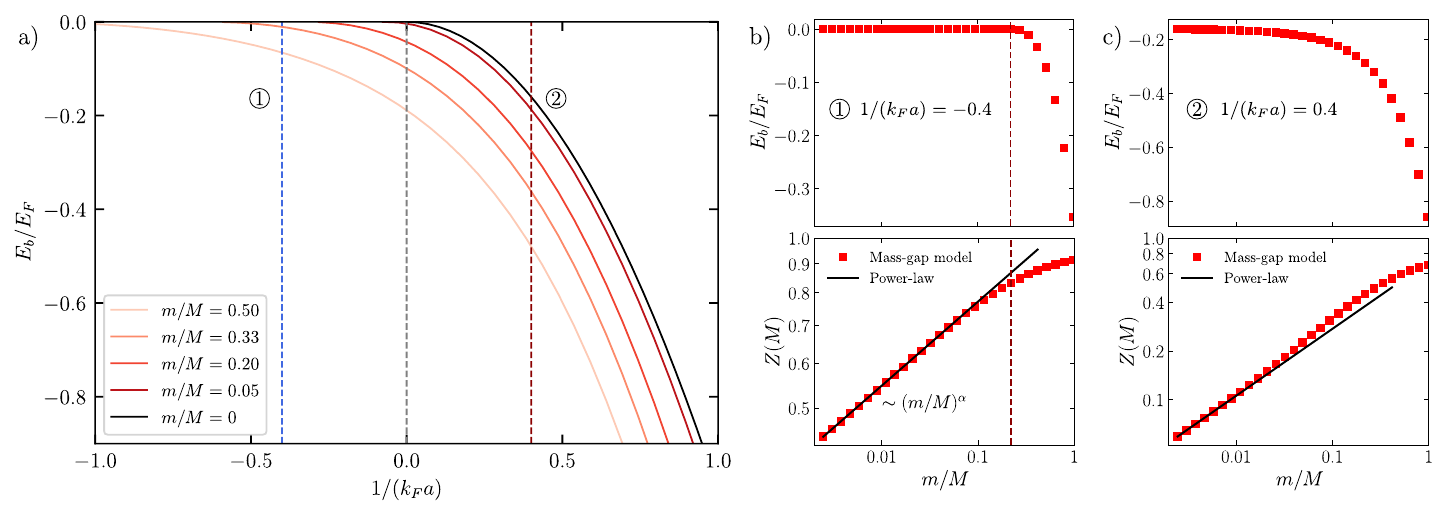}
    \caption{Impact of the bound state \circleletter{b} on the quasiparticle weight decay $Z(M)\sim (m/M)^{\alpha}$. a) The emergence of the bound state shifts toward $1/(k_Fa)\rightarrow 0$ as the impurity mass $M$ increases. b) For negative scattering lengths, the bound state disappears beyond a critical mass ratio and the power-law decay of the quasiparticle weight sets in. c) For positive scattering lengths, the bound state is always present and converges to the result for an infinitely heavy impurity.}
    \label{fig:bound-state-dependence}
\end{figure}

Finally, we analyze the quasiparticle weight $Z_{\text{pol}}=|\langle\mathrm{FS}|\text{pol}\rangle|^2$ of the attractive polaron as a function of mass ratio $M/m$, in the continuum limit $N\rightarrow\infty$, and provide numerical results for the power-law decay $Z(M)\sim (m/M)^{\alpha}$ across the entire phase diagram. We also discuss the role and impact of the bound state \circleletter{b} on the power-law decay.

As described in the main text, for a finite mass ratio and fermion density, the bound state \circleletter{b} already appears at negative scattering length $a<0$ and modifies the power-law decay of the quasiparticle weight $Z(M)$. However, as shown in Fig.~\ref{fig:bound-state-dependence}a), the appearance of the bound state shifts toward $1/(k_Fa)\rightarrow 0$ as the impurity becomes heavier. In the limit $m/M\rightarrow 0$, the bound state no longer exists at negative scattering lengths and recovers the analytical result of an infinitely heavy impurity, $E_b=-1/(2ma^2)$ for $a>0$. This implies that, for any negative scattering length $a<0$, the bound state \circleletter{b} disappears at a certain mass ratio and does not modify the power-law decay.

Strikingly, the mass-gap model predicts a critical mass ratio beyond which the power-law decay of $Z(M)$ sets in. As shown in Fig.~\ref{fig:bound-state-dependence}b), we find that, for a given interaction $1/(k_Fa)<0$, the onset of this decay occurs only after the bound state disappears (indicated by the vertical dashed line). For weaker interactions, this happens at smaller mass imbalances ($M/m>2$) compared to the larger mass imbalances ($M/m>10$) required for stronger interactions.

For positive scattering length $a>0$, the bound state \circleletter{b} slowly approaches a finite energy and its permanent presence modifies the power-law scaling even for very large mass ratios, see Fig.~\ref{fig:bound-state-dependence}c). Therefore, a possible power-law decay sets in only at much larger mass ratios.

In Fig.~\ref{fig:Z-factor-decay-all}, we show numerical results for the power-law decay of the quasiparticle weight $Z(M)\sim (m/M)^{\alpha}$ for negative and positive scattering lengths. For negative scattering lengths, the power-law exponent $\alpha$ agrees with the system-size scaling of an infinitely heavy impurity, $Z(N)\sim N^{-\alpha}$ with $\alpha=(\delta_F/\pi)^2$. For positive scattering lengths, the exponent deviates from that of an infinitely heavy impurity, $\alpha=(1+\delta_F/\pi)^2$~\cite{Knap:2012OC}, due to possible subleading contributions from the mass-dependent bound state.

\begin{figure}[b!]
	\centering
	\includegraphics[width=0.6\textwidth]{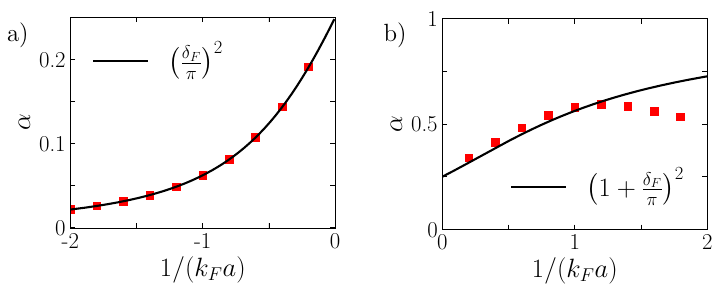}
	\caption{Numerical results for the power-law exponent $\alpha$ of the quasiparticle weight decay $Z(M)\sim (m/M)^{\alpha}$ (red symbols) for a) negative scattering lengths $a<0$ and b) positive scattering lengths $a>0$, in comparison to the analytic results for the power-law decay of an infinitely heavy impurity, $Z(N)\sim N^{-\alpha}$ (solid line).}
	\label{fig:Z-factor-decay-all}
\end{figure}

\end{widetext}
\end{document}